\documentclass[conference]{IEEEtran}
\usepackage[english]{babel}
\usepackage{amsmath}
\usepackage{amssymb}
\usepackage{amsthm}
\usepackage{booktabs} 

\usepackage{multirow}

\usepackage{dblfloatfix}

\usepackage{color}
\usepackage{diagbox}

\usepackage[T1]{fontenc}
\usepackage[export]{adjustbox} 

\usepackage{comment}
\usepackage{verbatim,epsfig,times,subfigure}

\usepackage{url}
\usepackage{hyperref}

\usepackage[colorinlistoftodos]{todonotes}

\makeatletter
\providecommand\@dotsep{5}
\renewcommand{\listoftodos}[1][\@todonotes@todolistname]{%
 \@starttoc{tdo}{#1}}
\makeatother

\newtheorem{definition}{{\bf \hspace{-0.18in} Definition}}
\newtheorem{assumption}{{\bf \hspace{-0.18in} Assumption}}
\newtheorem{theorem}{{\bf \hspace{-0.18in} Theorem}}
\newtheorem{corollary}{{\bf \hspace{-0.18in} Corollary}}
\newtheorem{lemma}{{\bf \hspace{-0.18in} Lemma}}

\def\done{\hspace*{\fill} \rule{1.8mm}{2.5mm}}

\begin{document}
\title{Zero-Rating and Net Neutrality: Who Wins, Who Loses?}




\author{\IEEEauthorblockN{ Niloofar Bayat}
\IEEEauthorblockA{
\textit{Columbia University}\\
niloofar.bayat@columbia.edu}
\and
\IEEEauthorblockN{Richard Ma}
\IEEEauthorblockA{
\textit{National University of Singapore}\\
tbma@comp.nus.edu.sg}
\and
\IEEEauthorblockN{Vishal Misra}
\IEEEauthorblockA{
\textit{Columbia University}\\
vishal.misra@columbia.edu}
\and
\IEEEauthorblockN{Dan Rubenstein}
\IEEEauthorblockA{\textit{Columbia University}\\
danielr@columbia.edu}
}

\maketitle

\begin{abstract}
An objective of network neutrality is that the design of regulations for the Internet will ensure that it remains a public, open platform where innovations can thrive. 
While there is broad agreement that preserving the content quality of service falls under the purview of net neutrality, the role of differential pricing, especially the practice of \emph
{zero-rating} remains controversial. Even though some countries (India, Canada) have banned zero-rating, others have either taken no stance or explicitly allowed it (South Africa, Kenya, U.S.). 

In this paper, we model zero-rating options available between Internet service providers (ISPs) and content providers (CPs)
 and use these models to better understand the conditions under which offering zero-rated services are preferred, and who specifically gains in utility. We develop a formulation in which providers' incomes vary, from low-income startups to high-income incumbents, and where their decisions to zero-rate are a variation of the traditional prisoner's dilemma game. 
 We find that if zero-rating is permitted, low-income CPs often lose utility,
 whereas high-income CPs often gain utility. We also study the competitiveness of the CP markets via the \emph{Herfindahl Index}. Our findings suggest that in most cases the introduction of zero-rating \emph{reduces} competitiveness.

\end{abstract}

\section{Introduction}\label{sec:intro}
Network neutrality, or simply net neutrality, is the principle that Internet service providers (ISPs) treat all data on the Internet equally, and do not discriminate or charge differently by user, content, website, platform, application, type of attached equipment, or method of communication \cite{gilroy2011access,wu2003network}. 
Net neutrality advocates claim that such discrimination weakens content providers' (CP) investment incentives since ISPs can expropriate some of the investment benefits \cite{pil2010net,gans2015weak}. However, net neutrality opponents argue that \emph{differential pricing}, which allows
ISPs to charge \emph{end users} different prices for data originating from different websites (CPs), triggers investments in broadband capacity and content innovation \cite{bourreau2015net}. Although by some definitions, net neutrality does not include issues involving pricing, it is worthwhile to study how differential pricing impacts the market and how data is ultimately treated, where net-neutrality unarguably centers (i.e., user content, etc.).


A commonly used practice of differential pricing is \emph{zero-rating}: a service where ISPs do not charge customers for bandwidth consumed by specific applications and services, while customers pay the bandwidth price for other used services. Today, zero-rating is used in practice by particular cellular network providers offering free access to selected online services~\cite{carrillo2015having}
. Proponents of zero-rating argue that offering some services for free increases customer satisfaction and online services usage, as they can access more data at a given cost \cite{West2015improve}. Critics argue that zero-rating leads to settings where sponsored data allows more financially prosperous CPs to pay for placement, which adversely affects smaller CPs who cannot afford the same luxury \cite{Ziegler14, Tonner16}. From this perspective, zero-rating may create artificial scarcity and jeopardize the achievement of the net neutrality rationale \cite{belli2017net}. 

Unfortunately, arguments to date, such as the above, are qualitative, and there is no formal model that would allow concerned parties to quantitatively ascertain how zero-rating ultimately impacts the marketplace. Thus, there is no way to quantify the extent of the win or loss that a heterogeneous mix of ISPs, CPs, and users experience based on enabling or disabling a zero-rated service.

In this paper, we formally model Internet settings in which zero-rating is a service offered by {\textbf ISPs} to the {\textbf CPs} who deliver their content to {\textbf users}, who are customers and obtain CPs' contents through their chosen ISPs. We perform $1$) a \textbf{macroscopic} analysis, i.e. zero-rating impacts on the competitiveness of the market as a whole, $2$) a \textbf{microscopic} analysis, i.e., zero-rating impacts on the behavior and decisions of individual ISPs and CPs. We use our models to quantitatively analyze how zero-rating impacts the market when ISPs, CPs, and consumers each choose options that maximize their individual rewards.

 Our model differentiates CPs in terms of their \emph{value}, i.e., how much revenue a CP makes per bandwidth unit used by their customers. 
 While incumbents typically have higher values, startups would have lower values, making less money per unit of bandwidth due to their smaller size and smaller market popularity. While our model and theoretical results are general for \emph{any} kind of differential pricing, in this paper we tailor our analysis explicitly to the zero-rating context, as it is the only prevalent real-world implementation of late.
 
 The new knowledge can help providers make informed zero-rating decisions and guide regulators to design better policies to address net neutrality issues from an interconnection context. Our contributions and conclusions are as follows.

\begin{itemize}
\item We consider both ISPs' and CPs' zero-rating decisions as a bargaining problem, analyze their strategic behavior, and introduce the concept of \emph{zero-rating equilibria}.

\item We identify the phenomenon of \emph{zero-rating pressure} where CPs only zero-rate because their competitors do, and find how it impacts CPs' decisions and their utilities.
\item We analyze the impact of zero-rating on the market of CPs both globally by analyzing the \emph{Herfindahl index} \cite{rhoades1993herfindahl}, and individually by analyzing CPs' utilities. The Herfindahl index analysis shows that zero-rating availability in a heterogeneous market of CPs increases market distortion by decreasing competition. Our utility analysis demonstrates that zero-rating will mostly negatively impact the utility of low-value CPs and positively impacts the utility of high-value CPs. 
\item We numerically explore the parameter space of our model and demonstrate the impact of zero-rating on market shares and profitability of the CPs under varying market conditions. 
\end{itemize}

The rest of the paper is organized as follows. In Section \ref{sec:mode} we build our choice model which takes ISPs and CPs as complementary services and characterizes their market shares (Equation \ref{equation:market_share}). In Section \ref{sec:zre} we build our utility model (Equation \ref{equation:utilities}
) under various zero-rating and market structures. Section \ref{sec:analysis} theoretically analyzes the Herfindahl index and utilities under zero-rating equilibria, and in Section \ref{sec:eval} we numerically and qualitatively measures the Herfindahl index and CPs' utilities after identifying zero-rating equilibria in a 
duopolistic market of ISPs/CPs. In Section \ref{sec:rel-work} we present the related work, and finally the paper is concluded in Section \ref{sec:con}.

\section{Model}\label{sec:mode}
We consider a setting with 3 types of players: \emph{user, ISP, and CP}. To receive content, a user must select one of $|\mathbb{M}|$ ISPs as their bandwidth provider and can choose from one or more of the $|\mathbb{N}|$ CPs from which the content is derived \footnote{Note that the ISP's side of the market can be easily extended to a model where users may have multiple ISPs at a time, similar to the way CP's side is modeled. However, since it is a realistic scenario where each customer has one ISP and multiple CPs, we set that as the basis of our model. }. For example, a user may select AT\&T as their ISP, and select Netflix, HBO, and Facebook as their CPs. Generally, a user pays its ISP for bandwidth, and the CP may also pay the ISP to deliver its content to users. Furthermore, a CP's income could come from a subscription fee the user pays them monthly, advertisement services, etc. In our analysis, we compute each CP's income based on the bandwidth unit of data consumed from it. For instance, if an average user consumes $10$ GB of data per month from a given CP, and the CP makes $\$10$ per month on average from each user, the CP's per-bandwidth unit income would be $1$ /GB. Such payments are the basis of a zero-rating service, based on which we develop our model through a sequence of steps explained in the following subsections.

\subsection{Dummy ISP and CP} \label{sec:auxiliary} 

In some cases, there exist users who may not choose any ISPs, and/or may not choose any CPs. For instance, users who select an ISP but no CP represent the ones who utilize Internet access but do not pay for any premium services like Netflix or HBO. Users who select CPs but no ISP are users who receive a premium service via another means, e.g., Netflix by DVD through mail. Users who select no ISP and no CP are those users who require neither of these services. Furthermore, some users may utilize multiple CPs at a time. In other words, each user may choose a combination of the existing CPs with a certain probability. 

To facilitate subsequent analysis, rather than having users who opt to select no ISP or no CP (or neither), we assume the existence of a \emph{dummy} CP and \emph{dummy} ISP. Users who would normally select no ISP (CP) can be mapped to a setting where they select the dummy ISP (CP) at no cost. Note that the dummy ISP carries no traffic and the dummy CP has no content to offer. 

 A user, therefore, has a choice of $|\mathcal{M}| = |\mathbb{M}|+1$ ISPs, including the dummy ISP, and can choose one of $|\mathcal{N}| = 2^{|\mathbb{N}|}$ possible subsets of CPs for their content, including the null set, which corresponds to the dummy CP. We refer to the subset of CPs chosen by the user as \emph{auxiliary CPs} and denote the set of ISPs and CPs (including dummies and auxiliaries) by $\mathcal M$ and $\mathcal N$, 
 respectively. Thus, a user always picks an ISP from $\mathcal M$ for her Internet access and one auxiliary CP from $\mathcal N$ for the content.
 
 \begin{figure}[t]
\centering
\hbox{\hspace{-1em}\includegraphics[width=4.5in, angle=0]{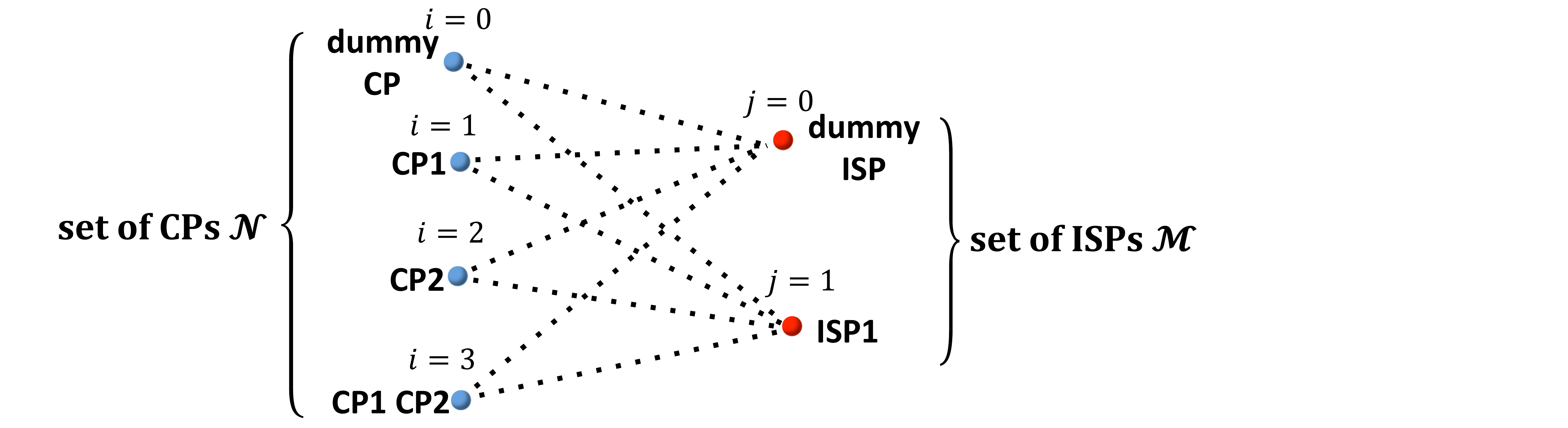}{}} 
\caption{The ISP-CP relations in a market of two real CPs and one real ISP, where we assume there also exist one dummy ISP, one dummy CP, and another auxiliary CP. Each user can choose one ISP $j\in \mathcal{M}$ and one CP $i \in \mathcal{N}$, and will be assigned to edge $(i,j)$ accordingly. 
}
\label{fig:theta}
\end{figure}

Figure \ref{fig:theta} illustrates the case where the market originally has $|\mathbb{N}| = 2$ CPs and $|\mathbb{M}| = 1$ ISP. We model the market by including the dummy ISP and auxiliary CPs, where we have $|\mathcal{N}| = 2^{|\mathbb{N}|} = 4$ and $|\mathcal{M}| = |\mathbb{M}| + 1 = 2$. Note that the relationship between CPs and ISPs is illustrated in a complete bipartite graph, where the users are assigned to different edges. More specifically, if a user accesses CP $1$ through ISP $1$, she will be assigned to the edge connecting $i=1$ to $j=1$. If a user chooses ISP $1$ as her provider and both CP $1$ and CP $2$, she will be assigned to the edge connecting $i = 3$ to $j = 1$. Each user can be assigned to one and only one edge, therefore, the number of customers assigned to different edges should sum up to the total market size.
 
 \subsection{Zero-Rating} \label{sec:zr}
 In a paradigm where zero-rating is permitted, an ISP and a CP may agree that instead of the user, the CP will cover the \emph{bandwidth} cost of the content viewed by the user, potentially at a price lower than what the user would pay. 
 A user's choice of ISPs and CPs could depend upon whether zero-rating is permitted and offered by an ISP-CP pair as a service.

We assume customers are mapped to an ISP-CP pair $(i,j)$ according to a pre-determined distribution. However, the distribution is affected by zero-rating choices: when ISP $j$ and CP $i$ decide to zero-rate, the probability of customers being assigned to them changes. We denote the zero-rating relationship between CP $i\in\mathcal N$ and ISP $j\in\mathcal M$ by $\theta_{ij}\in\{0,1\}$, where $\theta_{ij} = 1$ indicates zero-rating between $i$ and $j$ is established, otherwise $\theta_{ij} = 0$. 

Even though zero-rating does not apply to the dummy CP $i=0$ or ISP $j=0$, we always assume $\theta_{0j} = \theta_{i0} = 0;\ \forall i\in \mathcal{N},\ j\in \mathcal{M}$. In dummy providers, zero-rating is an abstract term, and as we will see in Section \ref{sec:user-model}, these providers are always assumed not to have any zero-ratings, and they capture some elastic users of the market in case no actual provider zero-rates, as well as their own sticky users. Different from dummy providers, zero-rating relation can be extended to auxiliary CPs. Since an auxiliary CP is a subset of actual CPs, it is assumed to have zero-rating relation with a given ISP $j$ if and only if all the CPs comprising it zero-rate with the ISP. For instance, in Figure \ref{fig:theta} the auxiliary CP $3$ is assumed to have a zero-rating relation with ISP $1$ if and only if both CP $1$ and CP $2$ have a zero-rating relation with ISP $1$, i.e., $\theta_{31} = 1 \iff \theta_{11}=1\ \&\ \theta_{21}=1$.

\subsection{Complementary Choices Model $(\mathcal{N},\mathcal{M})$}\label{sec:comp-model}
We denote the \emph{baseline market share} of ISP $j$ by $\psi_j\in(0,1]$, which captures the \emph{intrinsic} characteristics such as price and brand name, and models the market share of ISP $j$ when none of ISPs in the system have zero-rating relationships with the CPs. In that case, the percentage $\psi_j$ of each CP $i$'s users will choose ISP $j$ from the set $\mathcal M$ of ISPs.
In probabilistic choice models \cite{McFadden80}, $\psi_j$ can also be interpreted as the probability that any user chooses ISP $j$ where none of ISPs offer zero-rating, and we have $\sum_{j\in\mathcal M} \psi_j=1$. 
Furthermore, we denote the baseline market share of CP $i$ by $\phi_i\in(0,1]$, i.e., the market share of CP $i$ when none of CPs zero-rate with the ISPs, and we have $\sum_{i\in\mathcal N} \phi_i=1$. 

We define $\boldsymbol \psi \triangleq (\psi_{1},\cdots,\psi_{|\mathcal{M}|})^T$ and $\boldsymbol \phi \triangleq (\phi_{1},\cdots,\phi_{|\mathcal{N}|})^T$. Similarly, we have $\boldsymbol{\theta_i} \triangleq (\theta_{i1},\cdots,\theta_{iM})$ and $\boldsymbol{\vartheta_j}\triangleq (\theta_{1j},\cdots,\theta_{Nj})$ as the zero-rating profile of CP $i$ and ISP $j$, respectively. 
The zero-rating matrix of the whole system is defined as $\Theta \triangleq\left\{\theta_{ij}:i\in{\mathcal N},j\in{\mathcal M}\right\}$.
Given any zero-rating strategy $ \Theta$, we denote the strategies of all pairs of providers other than CP $i$ and ISP $j$ by $\Theta_{-ij}$.

We define $\alpha$ to be the fraction of elastic users of the market who choose among the CPs and ISPs with zero-rating relations, and if no such providers exist, these users would be distributed among all the providers. The rest of the users are distributed among CPs and ISPs merely based on their baseline market shares and independent of the zero-rating relations. These users are referred to as non-elastic, or sticky users, who comprise $1-\alpha$ fraction of the market. Note that the sticky users do not move among the providers if their zero-rating relations change, while elastic users do. 

\begin{table}[t]
\centering
\footnotesize
\begin{tabular}{c c} 
\toprule
parameter & description \\
\hline
$(i,j)$ & a pair of CP $i$ and ISP $j$ \\
\hline
$\mathcal{N}$, $\mathcal{M}$ & set of all CPs, ISPs (including auxiliary) \\
\hline
$\mathbb{N}$, $\mathbb{M}$ & set of actual CPs, ISPs \\
\hline
$\theta_{ij}$ & zero-rating relation between $(i,j)$ \\
\hline
 $X_{ij}$ & \#users of $(i,j);\ i\in\mathcal{N}, j \in \mathcal{M} $ \\
\hline 
 $\mathbb{X}_{ij}$ & the effective \#users of $(i,j);\ i\in\mathbb{N}, j \in \mathbb{M}$ \\
\hline
$\alpha$& user elasticity\\
\hline
$c$& bandwidth usage coefficient for non-zero-rated data \\
\hline
 $q_{i}$ & per bandwidth revenue of CP $i$; $ i\in\mathbb{N}$\\
\hline 
$p_j$ & per bandwidth price of ISP $j$; $ j \in \mathbb{M}$\\
\hline 
 $\phi_{i}$ & baseline market share of CP $i$; $ i\in\mathcal{N}$\\
\hline 
 $\psi_{j}$ & baseline market share of ISP $j$; $ j \in \mathcal{M}$\\
\hline 
 $\delta_{j}$ & price discount of ISP $j$; $ j \in \mathbb{M}$\\
\hline 
 $U_i, R_j$ & utility of CP $i$, ISP $j$; $ i\in\mathbb{N}, j \in \mathbb{M}$\\
\bottomrule 
\end{tabular}
\caption{Summary description of parameters.}
\label{Tab:parameters}
 \end{table}

In practice, users may choose services from constrained sets of CPs and ISPs. It might be because certain providers are not available to the users or cannot satisfy their requirements. In general, we denote a set of choice pairs by $\mathcal L$. This set of available choices is impacted by zero-rating relations in the system, i.e., $\Theta$. If CP $i$ stops the zero-rating relationship with ISP $j$ such that the bandwidth costs for their services on that ISP is no longer zero, their users might switch to an alternative ISP and/or CP in the market.

{\assumption The set $\mathcal L$ for sticky users is the entire choice set $\mathcal N \times M$. The set $\mathcal L$ for elastic users is any pair of CP $i$ and ISP $j$ who have zero-rating relations with one another, and if there is no such pair, it would be the entire choice set $\mathcal N \times M$.
\label{assumption:stickiness}}

Assumption \ref{assumption:stickiness} states that elastic users of the market only choose among pairs of CP-ISP which have zero-rating relations (if such pairs exist, otherwise, they choose among all CP-ISP pairs), since they want to maximize their surplus. However, since the sticky users in the market are not impacted by zero-rating relations, they are distributed among all ISPs and CPs based on their baseline market shares. In other words, they stay with their initial providers regardless of $\Theta$. 

Based on the baseline market shares of the providers, we make the following assumption on the users' choices.

{\assumption Given a nonempty set $\mathcal L$ of available choices, a user chooses a choice pair $(i,j)\in\mathcal L$ with probability
\begin{equation}
\mathbb{P}_{\mathcal L}\left\{(i,j)\right\} = \displaystyle\frac{\phi_{i}\psi_{j}}{\sum_{(n,m)\in\mathcal L}\phi_{n}\psi_{m}}
\label{equation:choice}
\end{equation}
\label{assumption:choice}}

Under Assumption \ref{assumption:choice}, if $\mathcal L$ equals the choice set $\mathcal N \times M$, the probability of choosing $(i,j)$
equals $\phi_{i}\psi_{j}$, which is consistent with our notion of \emph{baseline market shares}. Furthermore, by Luce's choice axiom \cite{Luce59}, the proportional form in (\ref{equation:choice}) is also necessary for guaranteeing an \emph{independence from irrelevant alternatives} (IIA) property: the probability of selecting one item over another from a pool of many items is not affected by the presence or absence of other items in the pool.

\subsection{User Model}\label{sec:user-model}
Our complementary choices model $(\mathcal N, M)$ can be entirely specified by a triple of vectors $(\boldsymbol{\phi},\boldsymbol{\psi},\boldsymbol{\Theta}$) and the scalar $\alpha$. We denote the total market size by $X$.
Based on Assumptions \ref{assumption:stickiness} and \ref{assumption:choice}, we characterize the number of users of $(i,j)$, denoted by $X_{ij}$, as a function of the zero-rating matrix $\Theta$ of the system as $X_{ij}(\Theta) = \rho_{ij}(\Theta)X$, where $\rho_{ij}$ is the closed form market share of the pair $(i,j)$ and we have:
\begin{multline}
\rho_{ij} (\Theta) =(\frac{\phi_i\psi_j \theta_{ij}}{\sum_{i'}\sum_{j'}\phi_{i'}\psi_{j'}\theta_{i'j'}}\alpha + \phi_i\psi_j(1-\alpha))\boldsymbol{1}_{\{\Theta \neq \boldsymbol{0}\}}\\ + (\phi_i\psi_j)\boldsymbol{1}_{\{\Theta = \boldsymbol{0}\}} \ \ \ \ \ \ \ \ \ \ \ \ \ \ \ \ \ \ \ \ \ \ \ \ \ \ \ \ \ \ \ \ 
\label{equation:market_share}
\end{multline}

Equation \ref{equation:market_share} derives the number of users $X_{ij}$ for any pair $(i,j)$ of complementary providers under the zero-rating profile $\Theta$. In particular, $X_{ij}$ can be represented by the number of users in the system $X$ multiplied by the closed form market share $\rho_{ij}(\Theta)$, which is a function of the zero-rating profile $\Theta$, the baseline market shares, and the elasticity of the users. If CP $i$ and ISP $j$ do not zero-rate, i.e., $\theta_{ij}=0$, the pair $(i,j)$ of providers can keep the proportion $1-\alpha$ of sticky users, who are distributed in the system based on baseline market shares. On the other hand, if CP $i$ zero-rates with ISP $j$, i.e., $\theta_{ij}=1$, not only they will keep the proportion $1-\alpha$ of their sticky users, but also the elastic users in the market would have incentives to choose them. Since elastic users of the market are distributed among the providers who offer zero-rating, $\alpha$ fraction of users would choose this pair with probability proportional to their baseline market shares. If none of the CPs and ISPs have zero-rating relations, all users are again distributed among the providers based on their baseline market shares.

{\lemma
Let ${\mathcal N}'\subseteq {\mathcal N}$ and ${\mathcal M}'\subseteq {\mathcal M}$. For any $n\notin{\mathcal N}$ and $m\notin{\mathcal M}$, let $\widetilde{\mathcal N} \triangleq {\mathcal N}\backslash{\mathcal N}'\cup\{n\}$ and $\widetilde{\mathcal M} \triangleq {\mathcal M}\backslash{\mathcal M}'\cup\{m\}$ denote the new sets of providers where the subsets ${\mathcal N}'$ and ${\mathcal M}'$ are replaced by the providers $n$ and $m$, respectively.
	Let $\widetilde{X}_{ij}$ denote the number of users of $(i,j)$ under $(\widetilde{\mathcal N},\widetilde{\mathcal M})$.
If $\boldsymbol{\theta_i} = \boldsymbol{\theta_n},$ $\ \forall i\in{\mathcal N}'$, $\boldsymbol{\vartheta_j} = \boldsymbol{\vartheta_m},$ $\ \forall j\in{\mathcal M}'$, and $(\phi_n,\psi_m) = \left(\sum_{i\in{\mathcal N}'} \phi_i, \sum_{j\in{\mathcal M}'} \psi_j\right)$, then
\[ \widetilde{X}_{nm} = \sum_{i\in{\mathcal N}'} \sum_{j\in{\mathcal M}'} X_{ij}; \quad \widetilde{X}_{ij} =X_{ij}, \ \forall i\neq n, j\neq m; \]
\[ \widetilde{X}_{nj} = \sum_{i\in{\mathcal N}'} X_{ij}, \ \forall j\neq m; \ \text{and} \ \ \widetilde{X}_{im} = \sum_{j\in{\mathcal M}'} X_{ij}, \ \forall i\neq n. \]
\label{lemma:additivity}}
Lemma \ref{lemma:additivity} states that if there exists multiple CPs (or ISPs) that use the same zero-rating profile, then they could be conceptually merged as a single CP (or ISP) without affecting the market shares of other providers.

When the users choose multiple CPs at the same time, they contribute to the revenues of all CPs they use, as well as the ISP they are assigned to. Let's assume that the set of CPs who are comprised from the actual CP $i$ is shown by $AUX(i)$. For instance, in Figure \ref{fig:theta}, we have: $AUX(1) = \{1,3\}$, and $AUX(2) = \{2,3\}$. Therefore, the effective number of users who utilize the pair of and actual CP $i$ and ISP $j$ can be computed from Equation \ref{equation:effective_user}.
\begin{equation}
\mathbb{X}_{ij} = \sum_{i\in AUX(i)}X_{ij} 
\label{equation:effective_user}
\end{equation}

{\corollary When all the providers $\mathcal{N}\times \mathcal{M} - \{(i,j)\}$ have fixed strategies, i.e., $\Theta_{-ij}$ is fixed, zero-rating $(i,j)$, i.e., changing $\theta_{ij}$ from $0$ to $1$, always helps CP $i$ to attract more customers. \label{lemma:customers}}

Based on Equation \ref{equation:market_share}, in case CP $i$ zero-rates with ISP $j$, since $\theta_{ij}=1$, as the first term of the equation is a positive value, it causes $X_{ij}$ to increase, which then increases $\mathbb{X}_{ij}$ based on Equation \ref{equation:effective_user}.

\section{Utility and Zero-rating Equilibria}\label{sec:zre}
Pricing takes various forms for the Internet in practice. Wireline ISPs often charge flat-rates \cite{anania97flat}, while tiered schemes have recently been adopted by major U.S. broadband providers such as Verizon \cite{Segall} and AT\&T \cite{Taylor}. Internet transit services, however, are usually charged based on peak rates, e.g., the 95th percentile measurement \cite{Stanojevic10}. 
Although data pricing may take various forms, we assume each ISP and CP choose whether or not to adopt zero-rating between each other for their customers. 

Although the data prices and values are assumed to be exogenous, each ISP $j$ has the option of charging a different data price $\delta_j p_j$ from CPs in case they have zero-rating relation, where $0<\delta_j \leq 1$ denotes the data price discount ISP $j$ offers to the CPs. If $\delta_j < 1$, CPs can purchase ISP $j$'s bandwidth as a zero-rated service, with a lower price than the users can directly purchase it. Even though in this case ISP $j$ loses per bandwidth unit income, its total revenue may increase since it could attract a higher number of customers. However, a low value of $\delta_j$ could harm ISP $j$'s total revenue. 
We define $\boldsymbol \delta \triangleq (\delta_{1},\cdots,\delta_{|\mathcal{M}|})^T$ to denote the entire ISP discount profile of the market.

In this section, we analyze ISPs' and CPs' zero-rating decisions. Note that although the auxiliary providers do not make independent zero-rating decisions, we account for their users in our evaluations because their users contribute to the users of actual providers (Equation \ref{equation:effective_user}), and therefore to their utilities. 
However, the users of dummy providers do not generate any utilities. We first introduce the following assumption to compute the utility model of the actual providers in the market.

{\assumption
The revenue of ISP $j\in \mathbb{M}$ from the market of CPs is equal to the summation of revenues each CP $i$'s user brings to $j$ for all $i\in \mathbb{N}$. Likewise, the utility of each actual CP $i\in \mathbb{N}$ is equal to the summation of utility each ISP $j$'s user brings to $i$ for all $j\in \mathbb{M}$. \label{assumption:ISP-revenue}
}

Note that even though the utility is a general term, it can also model the benefit a player gains in an abstract form. Since unlike ISPs, each CP's income has an indirect relationship with the bandwidth usage, we use the term \emph{utility} to model its decision-making process. Whereas to avoid confusion, we use the term \emph{revenue} to address the same thing for ISPs.

{\assumption
When zero-rating is provided for a pair of CP $i$ and ISP $j$, i.e., $\theta_{ij}=1$, since users would not pay the bandwidth price, their average bandwidth usage may increase by a factor of $1/c$, where $0< c\leq1$.
\label{assumption:bandwidth_usage}}

When the bandwidth usage increases 
in case of zero-rating, the utilities of CPs and the revenues of ISPs who zero-rate will also increase since they are a function of per-bandwidth unit prices. To model this phenomenon, instead of assuming the utilities and revenues of providers who zero-rate increase by a factor $1/c\geq 1$, for simplicity we assume if the providers cancel their zero-rating, their utilities and revenues decreases by a factor of $c$, where $0\leq c\leq1$ and it is called \emph{bandwidth usage coefficient}. Note that our model is not designed to capture bandwidth saturation for the ISPs, assuming ISPs to be smart agents with mechanisms to provide the bandwidth requested by the users, and in case of zero-rating the CPs will pay for ISPs' bandwidth.

Using Assumptions \ref{assumption:ISP-revenue} and \ref{assumption:bandwidth_usage}, and given any zero-rating strategy profile $\Theta$, we denote the utility of CP $i$ by $U_i(\Theta)$ and the revenue of ISP $j$ by $R_j(\Theta)$ and define them as
\begin{multline}
U_i(\Theta)\triangleq\sum_{j\in \mathbb{M}} U_{i}^j(\Theta) \ \ \ \text{and} \ \ \ R_j(\Theta)\triangleq \sum_{i\in\mathbb{N}}R_j^i(\Theta),\\
\text{where} \quad U_{i}^j(\Theta) \triangleq
\begin{cases}
q_i \mathbb{X}_{ij}(\Theta).c & \text{if} \quad \theta_{ij}=0,\\
(q_i - \delta_j p_j) \mathbb{X}_{ij}(\Theta) & \text{if} \quad \theta_{ij}=1.
\end{cases},\ \ \ \ \ \ \ \ \ \ \ \ \ \\
\text{and}\ \ \ \quad R_{j}^i(\Theta) \triangleq
\begin{cases}
p_j \mathbb{X}_{ij}(\Theta).c & \text{if} \quad \theta_{ij}=0,\\
\delta_jp_j \mathbb{X}_{ij}(\Theta) & \text{if} \quad \theta_{ij}=1.
\end{cases}\ \ \ \ \ \ \ \ \ \ \ \ \ \ \ \ \ \ \ \ \ \ 
\label{equation:utilities}
\end{multline}
Each CP $i$'s utility is the sum over the utilities $U_{i}^j$ generated from each ISP $j$, which equals to the effective number of users $\mathbb{X}_{ij}(\Theta)$ multiplied by either its profit margin $q_i - \delta_jp_j$ if zero-rating is provided, or its original value $q_i$ otherwise. Similarly, each ISP $j$'s revenue is the sum over the revenues $R_{j}^i$ generated from each CP $i$, which equals the effective number of users $\mathbb{X}_{ij}(\Theta)$ multiplied by either its price $p_j$ if zero-rating is not provided, or by the weighted value of $\delta_j p_j$ otherwise. Note that the utility functions of auxiliary CPs are not well-defined since their users are being counted as effective users of the actual CPs they encompass, and the users of dummy providers do not generate any utility.

CPs' zero-rating decisions depend on the prices imposed by ISPs, i.e., $\delta_j p_j$ for ISP $j$. ISPs' decisions depend on the revenue they receive from CPs via zero-rating compared to what they would receive from users directly. Equation \ref{equation:market_share} and \ref{assumption:ISP-revenue} show that CP $i$'s utility $U_i(\Theta)$ depends not only on its own strategy $\theta_{ij}$, but also on all other CPs' and ISPs' relations $\Theta_{-ij}$.
We assume that given the price profile $\boldsymbol p$, ISPs make simultaneous zero-rating offers with deciding a discount profile $\boldsymbol \delta$ to maximize their revenues, and CPs make simultaneous decisions whether or not to adopt them. We define a zero-rating equilibrium as follows:

{\definition[Zero-rating Equilibrium]
In a market of ISPs and CPs, given fixed discount and price profiles, a zero-rating strategy profile is a \emph{zero-rating equilibrium} (ZRE) if and only if 1) given a zero-rating strategy $\Theta$ chosen by ISPs, neither of CPs would gain by unilaterally deviating from $\Theta$ 2) given a zero-rating strategy $\Theta$ chosen by CPs, neither of ISPs would gain by unilaterally deviating from $\Theta$.
\label{def:zero-rating-equilibrium}}

Based on Definition \ref{def:zero-rating-equilibrium}, if $\Theta$ is a ZRE, for each actual CP $i$ and ISP $j$ we have $U_i(\theta_{ij};{\Theta}_{-ij}) \geq U_i(\bar\theta_{ij};{\Theta}_{-ij})$ and $R_j(\theta_{ij};{\Theta}_{-ij}) \geq R_j(\bar\theta_{ij};{\Theta}_{-ij})$.

ZRE is a specific kind of Nash equilibrium \cite{Nash50}, where there exist two groups of inter-dependent players. Since zero-rating is a bilateral contract between ISPs and CPs, the zero-rating decision which is affected by the entire market resembles a bargaining problem. For instance, given a pair of ISP-CP, the CP (ISP) does not have the option of establishing zero-rating if the ISP (CP) is not willing to zero-rate. Therefore, 
we use the term ZRE to avoid confusion. ZRE is evaluated for a \emph{pure strategy} game since mixed strategy decisions between CP $i$ and ISP $j$ to zero-rate do not apply to real-world scenarios, and users need deterministic knowledge on which ISP-CPs offer zero-rating. 

In some ISP prices, although ZRE is where a CP $i$ zero-rates, it may face a utility drop compared to the case where no zero-rating is allowed in the market, i.e., $\Theta = \boldsymbol{0}$. In this case, if CP $i$ deviates, it loses customers to the CPs who zero-rate and its utility further drops. This scenario resembles \emph{prisoner's dilemma} paradox in \cite{nowak1993strategy}, where each player chooses to protect themselves at the expense of the other participant and as a result, the optimal outcome will not be produced. However, since the market of CPs is mostly heterogeneous, this scenario mainly harms the low-value CPs rather than high-value ones as we see in Section \ref{sec:util}. We define \emph{zero-rating pressure} to address this phenomenon.

{\definition[Zero-rating Pressure] Zero-rating pressure happens when a CP decides to zero-rate to avoid losing customers; only because another CP is zero-rating with the same or different ISP, and if the latter cancels its zero-rating relation, the former would not zero-rate.\label{def:zr-pressure}}

Suppose there are two heterogeneous CPs in the market, and in ZRE the CP with a lower value zero-rates with an ISP. In that case, its utility shall improve compared to when it does not zero-rate. If zero-rating increases low-value CP's utility, the same must be true for the high-value CP as well. We introduce Lemma \ref{lemma:high_value} to generalize this case. 

{\lemma In a market $\mathbb{N}$ of content providers with different values, suppose $i, i' \in \mathbb{N}$ and $q_i < q_{i'}$. The zero-rating strategy that CP $i$ zero-rates with ISP $j$ while CP $i'$ does not is never a ZRE. \label{lemma:high_value}}

A more rigorous proof of Lemma \ref{lemma:high_value} is provided in Appendix I, which will be used in the next section to analyze how zero-rating impacts the market as a whole, and its components as individuals. 




\section{Analysis} \label{sec:analysis}
In this section, we represent a macroscopic and microscopic analysis of the impact of zero-rating on the market of CPs. For the former, we evaluate the Herfindahl index \cite{rhoades1993herfindahl} of the market, which is a proxy of competitiveness in the entire market and looks at the CPs as a whole. For the latter, we look into individual CPs' utilities. 

\subsection{Herfindahl Index Analysis} \label{sec:Herf}
To show the impact of zero-rating on the market, we compute the \emph{Herfindahl index} \cite{rhoades1993herfindahl}, also known as \emph{Herfindahl-Hirschman index} or \emph{HHI}, for different ISP prices among CPs. This index is calculated as the sum of squares over the market shares of all firms in the market, and since it accounts for the number of firms and concentration, it is an indicator of competition among firms. When this index grows to $1$, the market moves from a collaborative state to a monopolistic content provider, i.e., the competition decreases. Lack of competition in the market causes market distortion and significant welfare loss due to monopoly \cite{cowling1978social}. HHI increases of over 0.01 generally provoke scrutiny, although this varies from case to case \cite{HHI_scrutiny}. 

In this section, we analyze a market of CPs with different values and show how the availability of zero-rating would impact the Herfindahl index in the market. The analysis of this section is based on the user model in Equation \ref{equation:market_share}, and since we draw the conclusions theoretically, they are general to our model and are independent of parameter choices. The detailed proof of the lemmas in this section are present in Appendix I.

{\lemma In the market $\mathbb{N}$ of content providers, the Herfindahl index increases when the variance of content providers' market shares
increases. \label{lemma:HHI}}

Based on Lemma \ref{lemma:HHI}, the more different the market shares of CPs are, the higher the Herfindahl index would be. Intuitively, a high variance between the market shares indicates that the market is moving towards a monopoly, where the increase in the Herfindahl index confirms that as well. 

{\lemma The Herfindahl index is the same if none of the CPs zero-rate versus if every CP in the market zero-rates. \label{lemma:HHI_ZR_noZR}}

 The amount of consumption may increase in case every CP zero-rates in the market compared to when no one zero-rates. Furthermore, if everyone zero-rates the elastic users with dummy CP would be utilizing an actual CP. However, the relative market share of CPs remains unchanged (see Appendix I for proof). Therefore, the Herfindahl index stays the same in both cases. 

{\theorem In a market $\mathbb{N}$ of content providers with values $\{q_1, q_2, ..., q_{|\mathbb{N}|} \}$, suppose $q_1 \leq q_2 \leq ... \leq q_{|\mathbb{N}|}$, and the content providers with higher values also have higher baseline market shares, i.e., $\phi_1 \leq \phi_2 \leq ... \leq \phi_{|\mathbb{N}|}$. If at least one of these inequalities is strict, zero-rating options in the market will cause the Herfindahl index to be non-decreasing in all possible ZRE. \label{theorem:HHI_q_phi}}

Based on Lemma \ref{lemma:HHI_ZR_noZR}, when zero-rating is available in the market, in case ZRE consists of either every CP or no CP zero-rate, the Herfindahl index stays the same. In other ZRE cases, if CPs with higher values and higher baseline market shares afford more zero-ratings than their low-value opponents, the Herfindahl index will increase (proof in Appendix I). This could represent the case where startups with low incomes and low initial baseline market shares join the market of CPs. The increase in the Herfindahl index implies that the market moves toward monopoly, where the startups would not survive.

The correctness of Theorem \ref{theorem:HHI_q_phi} is based on Lemma \ref{lemma:HHI_ZR_noZR} which uses the relative market shares (see Appendix I). However, since the same conclusion cannot be driven for CPs' utilities as the actual number of users matters rather than the relative number of users, in the next section, we numerically analyze the utilities of CPs and observe the impact of different parameters on ZRE.

\subsection{Utility Analysis} \label{sec:util}
Computing the utility for each content provider requires prior knowledge of the ZRE strategies. Note that for a two-player game (or more), neither existence nor uniqueness of Nash equilibria could be guaranteed; it is $\mathcal{NP}$-complete to determine whether the Nash equilibria with certain natural properties exist \cite{gilboa1989nash} and it is $\#\mathcal{P}$-hard to count the Nash equilibria \cite{conitzer2008new}. However, in a heterogeneous market of CPs with different values, based on Lemma \ref{lemma:high_value}, there are a limited number of zero-rating strategies that could become an equilibrium, i.e., the case where a low-value CP zero-rates with an ISP, while another CP with higher value does not, is never an equilibrium while the opposite can be. We focus on how zero-rating impacts CPs with different values, where incumbents and startups could be thought of as CPs with high and low values, respectively. We introduce Theorem \ref{theorem:high_value_CP_utility} to address the special case where the low-value CP cannot afford to zero-rate in the equilibria, while the high-value CP can. 

{\theorem \label{theorem:high_value_CP_utility} In a market of CPs with $|\mathbb{N}|\geq 2$, let CP $1$ have a lower value than CP $2$. If the low-value CP does not zero-rate with any ISP ($\theta_{1j} = 0\ \forall j\in \mathbb{M}$) while the high-value CP does ($\theta_{2j} = 1\ \exists j\in \mathbb{M}$), and the zero-rating strategy profile for the rest of CPs (other than CP $1$ and CP $2$) does not change, the utility of low-value CP $1$ will always decrease compared with when zero-rating is not available, while the utility of high-value CP $2$ increases or remains the same.}

Since utility analysis, in general, depends on the exact zero-rating strategies of the market after equilibria, which are not possible to be determined in a closed-form formula, we perform numerical evaluations to illustrate the impact of zero-rating on the CPs in next section.

\section{Evaluation and Results}\label{sec:eval}
In this section, we use our model to evaluate and analyze the zero-rating equilibria for a market with complementary duopoly, i.e., $|\mathbb{M}|=|\mathbb{N}|=2$, where two CPs and two ISPs compete on both sides of the market. We assume that we have the regulation of \emph{weak content provider net neutrality} \cite{gans2015weak}, meaning that each ISP charges the same price from every CP. For the simplicity of the result illustration, we also assume that ISPs' price discount profile $\boldsymbol{\delta} = \boldsymbol{1}$. Therefore, the ISPs charge the same price from the CPs in case of zero-rating as they would charge the users otherwise. In Appendix II, we have analyzed a case where ISPs get to decide on the discount profile $\boldsymbol{\delta}$ and have shown how it would impact the market under ZRE. 

As Lemma \ref{lemma:additivity} shows that providers with similar prices and zero-rating strategies can be merged, a duopolistic model provides a first-order approximation of market competitions from a provider's perspective such that all its competitors are considered as an aggregated provider that captures the remaining market share. We assume the elasticity of the users, baseline market shares, and prices are exogenous. Note that this evaluation can be extended to different parameter choices and is not prone to parameter selection. However, to determine these parameters in a real-world market, the reader may refer to some previous work that study the impact of zero-rating on mobile Internet usage and have been reviewed in Section \ref{sec:rel-work}. We use a general model and synthetically set the parameters to show how the market behaves under different conditions, and we show parameter selection {\bf does not qualitatively change} our results. All prices and revenues in our evaluation are normalized to $1$ and are not intended to reflect \emph{absolute} real-world values, rather the relative differences between ISPs and CPs. 

We compare two different hypothetical markets, one where zero-rating options are not allowed, the other one where zero-rating options are allowed and available, and ISPs and CPs decide on their zero-rating strategies where the market could reach the equilibria. If {\bf multiple ZRE} exist, we make an arbitrary choice to illustrate only one ZRE with the maximum number of zero-rated relations, and in case of a tie, the high value CP (in our case CP $2$) and ISP $2$ will have the maximum number of zero-rated relations. If {\bf no ZRE} exists, we assume the market-share and utility of CPs remain the same as if there are no zero-rating options allowed in the market. We show how the zero-rating strategies are made and hence how HHI and the utilities of CPs are different in these two markets.

\subsubsection{Benchmark Scenario}
We use a benchmark scenario in which $c = 0.5$, $\alpha = 0.5$, and $\boldsymbol{\delta} = \boldsymbol{1}$. We also have $\boldsymbol{\phi} = (0.1, 0.4, 0.4, 0.1)$, and $\boldsymbol \psi = (0.2, 0.4,0.4)$, where assuming the vector indices start from $0$, $\phi_0$ is the baseline market share of dummy CP, i.e., the fraction of users who do not use any CPs, $\phi_3$ is the baseline fraction of customers who use \emph{both} CP $1$ and CP $2$, and $\phi_1$ and $\phi_2$ are the baseline market shares of CP $1$ and CP $2$, respectively. Similarly, $\psi_0$ is the baseline market share of dummy ISP, and $\psi_1,$ and $\psi_2$ are the baseline market shares of ISP $1$ and ISP $2$, respectively. Without loss of generality, we assume $q_1 \leq q_2$ and normalize the prices such that $\boldsymbol{q}\leq \boldsymbol{1}$ and $\boldsymbol{p}\leq \boldsymbol{1}$.

\begin{figure}[t]
\centering
\includegraphics[width=1.35in, angle=0]{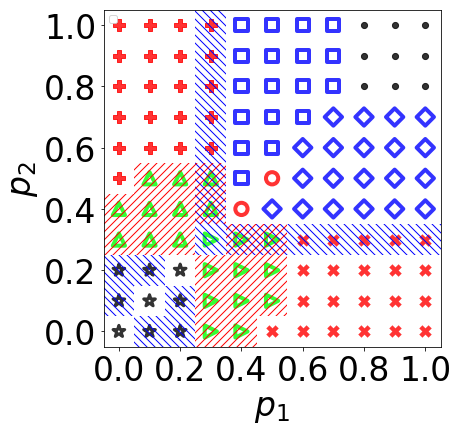} \raisebox{2mm}{\includegraphics[width=1.6in, angle=0]{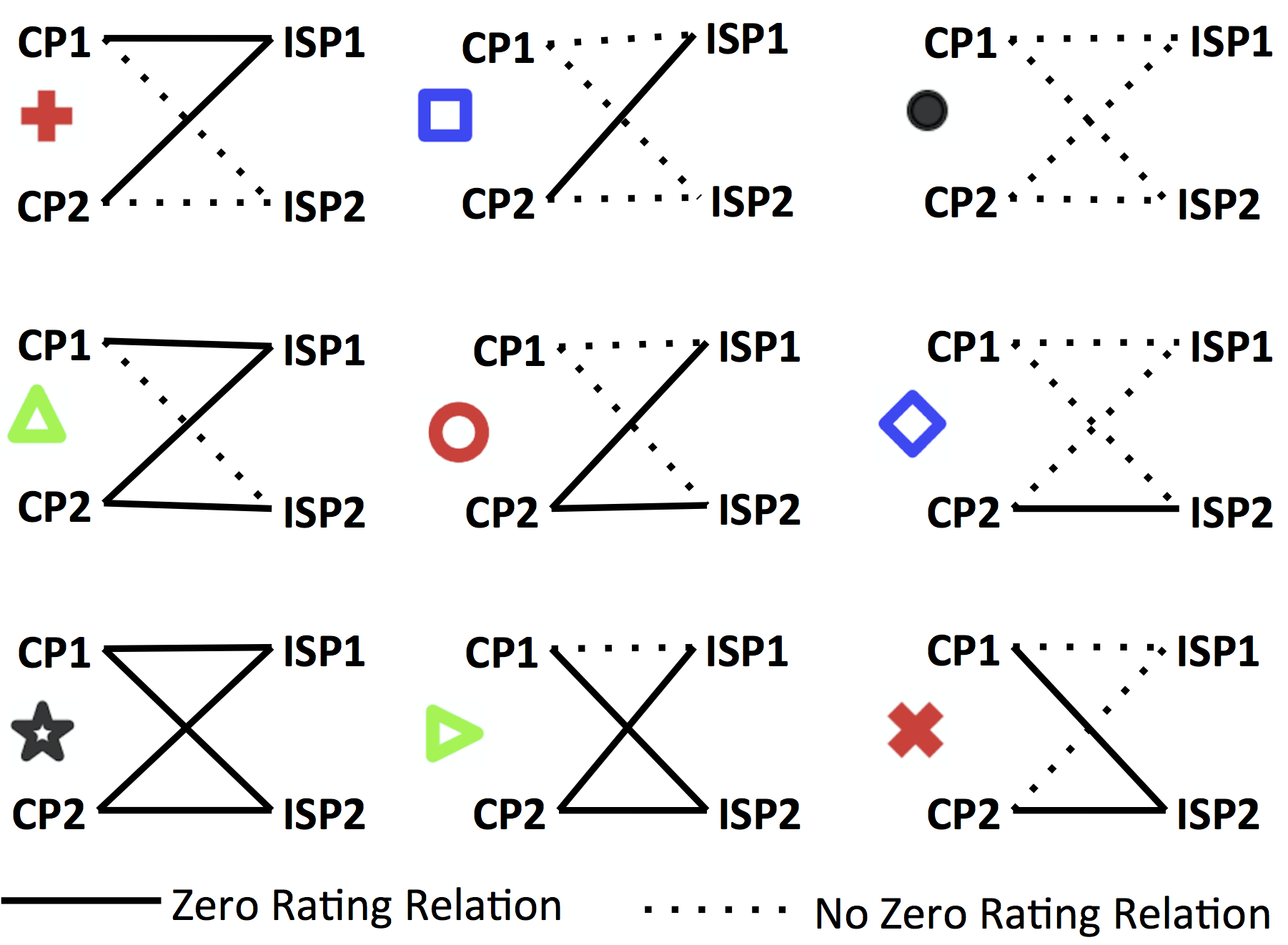}}
\caption{ZRE map under complementary duopoly with $\alpha = 0.5$, $c = 0.5$, $\boldsymbol{\delta} = (1.0, 1.0)$, $\boldsymbol{\phi} = (0.1, 0.4, 0.4, 0.1)$, and $\boldsymbol \psi = (0.2, 0.4,0.4)$. Shaded areas in blue ($\backslash$) and red ($/$) represent zero-rating pressure for CP $1$ and CP $2$, respectively.}
\label{fig:com_duopoly}
\end{figure}


Figure \ref{fig:com_duopoly} visualizes the ZRE when ISPs' prices $p_1$ and $p_2$ vary along the x- and y-axis, respectively. Based on Lemma \ref{lemma:high_value} as $q_2>q_1$, $9$ of the $16$ possible zero-rating profiles could become ZRE under various ISP prices, which are shown in the right sub-figure as legends. The case where ISP $j$ 
has the price of $0$ can represent when it offers an unlimited plan, and we assume in that case it is always zero-rating with all the CPs in the market. Intuitively, when ISP prices are low, both CPs are willing to zero-rate; but when the prices are high, neither CP is willing to do so. 
Low-value CP $1$ is generally more susceptible to ISP price changes; we observe that under any fixed price $p_{\bar j}$ as $p_j$ increases, CP $1$ first cancels its zero-rating with ISP $j$, followed by CP $2$.

Figure \ref{fig:com_duopoly} also illustrates the regions of zero-rating pressure for CPs. The shaded blue regions demonstrate when CP $1$ zero-rates under pressure, and the shaded red regions demonstrate when CP $2$ does so. Zero-rating pressure for high-value CP happens when the low-value CP establishes zero-rating with the cheaper ISP, the high-value CP as a response establishes a zero-rating relation with the more expensive ISP to maintain its customers. However, any zero-rating of the high-value CP can cause zero-rating pressure for the low-value CP, if it is not originally willing to zero-rate. 

\begin{figure*}[t]
\centering
\subfigure[CP $1$'s utility changes]{\includegraphics[width=2in, angle=0]{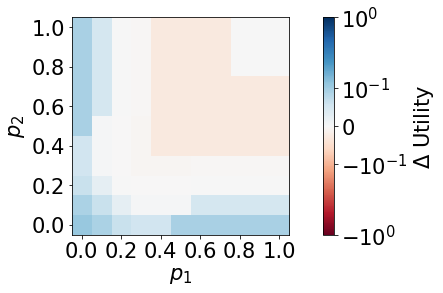}} 
\hspace*{-0.65in} 
\subfigure[CP $2$'s utility changes]{\includegraphics[width=2in, angle=0]{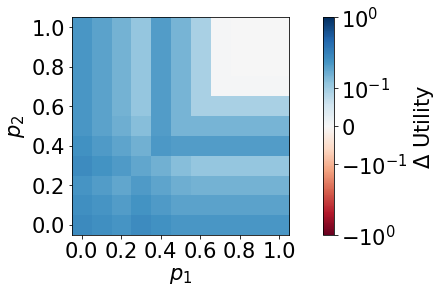}} 
\hspace*{-0.65in} 
\subfigure[HHI changes]{\includegraphics[width=2in, angle=0]{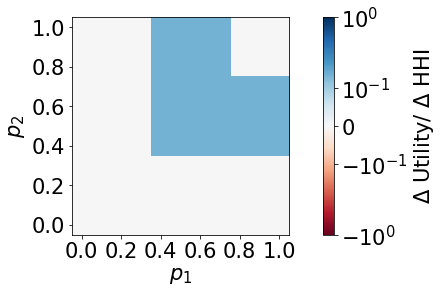}} 
\caption{the differences in CPs' utilities when zero-rating is available and the market reaches equilibria, minus when zero-rating is not available. We have: $\alpha=0.5$, $c=0.5$, $\boldsymbol{\phi} = (0.1, 0.4, 0.4, 0.1)$, $\boldsymbol{\psi} = (0.2, 0.4, 0.4)$, $\boldsymbol{\delta} = (1.0, 1.0)$ and $\boldsymbol{q}=(0.4,1.0)$.}
\label{fig:CP_utility_baseline}
\end{figure*}



 Figure \ref{fig:CP_utility_baseline} visualizes the impact of zero-rating on the CPs' utilities and the market HHI. We observe that in an imbalanced market of CPs, zero-rating \emph{usually} decreases the utility of low-value CP $1$, but increases the utility of high-value CP $2$. 
 Based on Lemma \ref{lemma:high_value} (and Figure \ref{fig:com_duopoly}), high-value CP $2$ always can afford more zero-rating relations than low-value CP $1$. Hence its utility mostly increases as it attracts more elastic users of the market. This figure also confirms Theorem \ref{theorem:high_value_CP_utility}, where in case CP $1$ does not have any zero-rating relations while CP $2$ does, CP $1$'s utility always decreases while CP $2$'s utility does not. In Figure \ref{fig:CP_utility_baseline}(c), we observe how HHI changes after ZRE as opposed to when it is not allowed, which is always non-decreasing in accordance to Theorem \ref{theorem:HHI_q_phi}. 

In the following subsections, with our focus on the market of CPs, we show how the changes in each parameter impact ZRE strategies, CPs' utilities, and HHI.

\subsubsection{Impact of Bandwidth Usage Coefficient}
To show how a decrease/increase of bandwidth usage during zero-rating impacts the market, we have varied the parameter $c$ and observed the equilibria, utilities, and HHI in the market. Since our benchmark analysis in Figure \ref{fig:com_duopoly} has $c = 0.5$, we test our model with both a higher and a lower value of $c$.

\begin{figure*}[t]
\centering
\subfigure[ZRE, $c=0.2$]{\includegraphics[width=1.35in, angle=0]{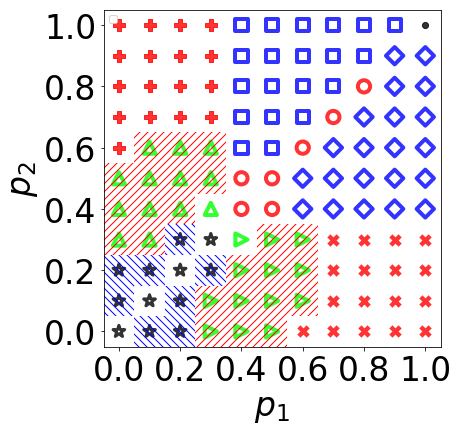}}
\hspace*{0.15in}
\subfigure[CP $1$'s utility changes]{\includegraphics[width=2in, angle=0]{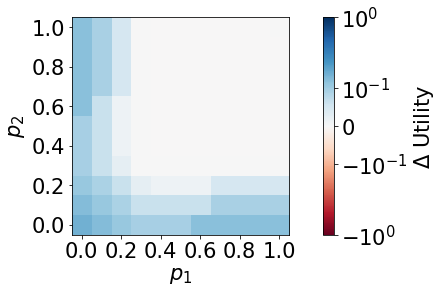}} 
\hspace*{-0.65in}
\subfigure[CP $2$'s utility changes]{\includegraphics[width=2in, angle=0]{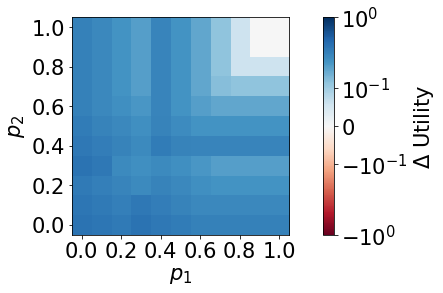}} 
\hspace*{-0.65in}
\subfigure[HHI changes]{\includegraphics[width=2in, angle=0]{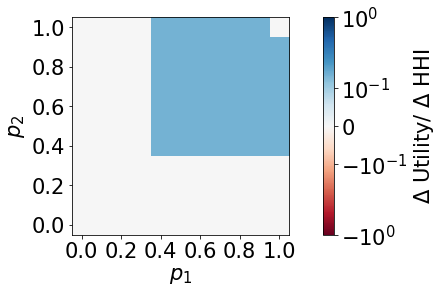}} 

\subfigure[ZRE, $c=0.8$]{\includegraphics[width=1.35in, angle=0]{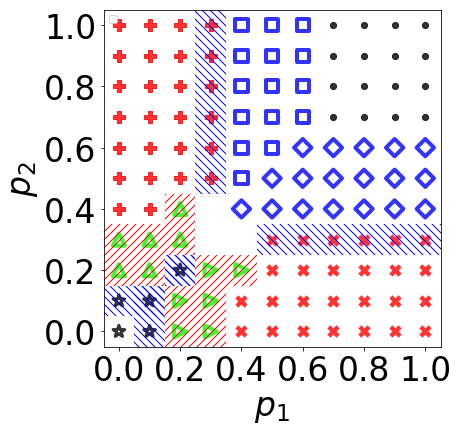}}
\hspace*{0.15in}
\subfigure[CP $1$'s utility changes]{\includegraphics[width=2in, angle=0]{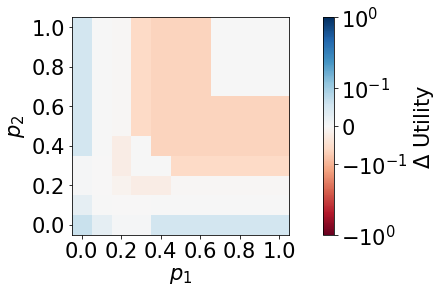}} 
\hspace*{-0.65in}
\subfigure[CP $2$'s utility changes]{\includegraphics[width=2in, angle=0]{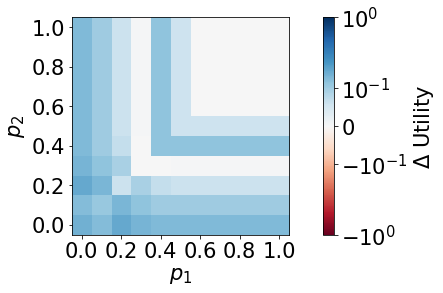}} 
\hspace*{-0.65in}
\subfigure[HHI changes]{\includegraphics[width=2in, angle=0]{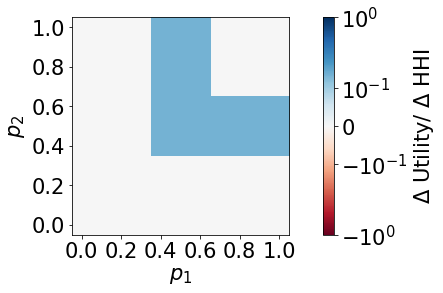}} 
\caption{(a,d) ZRE under complementary duopoly. (b,c,e,f) The differences in CPs' utilities when zero-rating is available and the market reaches equilibria, minus when zero-rating is not available. 
We have: $\alpha=0.5$, $\boldsymbol{\phi} = (0.1, 0.4, 0.4, 0.1)$, $\boldsymbol{\psi} = (0.2, 0.4, 0.4)$, $\boldsymbol{\delta} = (1.0, 1.0)$ and $\boldsymbol{q}=(0.4,1.0)$.}
\label{fig:impact_c}
\end{figure*}

Figure \ref{fig:impact_c}(a, e) illustrates the impact of $c$ on ZRE. When $c$ decreases from $0.5$ in figure \ref{fig:com_duopoly} to $0.2$ in Figure \ref{fig:impact_c}(a), the number of zero-rated relations increases as well since canceling zero-rating causes less bandwidth usage which produces less income for both CPs and ISPs. However, when $c = 0.8$ in Figure \ref{fig:impact_c}(e), canceling zero-rating would not decrease the bandwidth usage as much as when $c = 0.5$. Hence, the number of zero-rated relations decreases. The regions of zero-rating pressure for low-value CP $1$ also increase since as CP $1$'s incentives to zero-rate decrease, it is not willing to zero-rate unless its competitor is doing so, which is the definition of zero-rating pressure. Note that in Figure \ref{fig:impact_c}(e), there is no ZRE when $\boldsymbol{p}\in\{(0.3,0.3),(0.3,0.4),(04,0.3)\}$. In these regions, when CP $2$ zero-rates with an ISP, the best decision for CP $1$ would be to zero-rate as well. But if CP $1$ zero-rates, CP $2$ would cancel its zero-rating. As a response, CP $1$ would cancel its zero-rating as well, for which the best decision of CP $2$ would be to zero-rate. Therefore, there is an infinite loop where the market does not reach to any ZRE. Note that utility and HHI changes in these regions are assumed to be zero. 

Figure \ref{fig:impact_c} also depicts the utility and HHI changes corresponding to \ref{fig:impact_c}(a, d). We observe that when $c = 0.2$, both CPs may experience an increase in their utilities, which is more major for high-value CP $2$. Note that in regions where CP $2$ is the only one who zero-rates, as shown in Theorem \ref{theorem:high_value_CP_utility}, CP $1$'s utility decreases, although it might not be observable in the graphs. However, when $c = 0.8$, low-value CP $1$ experiences a more major utility loss, whereas the utility of CP $2$ may slightly increase or decrease. We also observe that HHI in both cases increases as expected by Theorem \ref{theorem:HHI_q_phi}. 

Note that when CP $1$ does not zero-rate but CP $2$ does, the utility difference between ZRE and when no zero-rating is allowed is smaller for CP $1$ when $c = 0.8$ compared to the benchmark ($c= 0.5$) and $c = 0.2$. When CP $1$ does not zero-rate but CP $2$ does, for all values of $c$ the utility difference for CP $1$ is negative. However, based on Equation \ref{equation:utilities}, since the utility of CP $1$ with no zero-ratings is multiplied by $c$, the larger $c$ is, the smaller this negative number would be after multiplication. 

\emph{Takeaways:} 
When the bandwidth usage coefficient ($c$) is small, both CPs experience utility gain after zero-rating, where this gain is higher for the high-value CP. When $c$ is in the mid-range ($c = 0.5$), low-value CP primarily experiences utility loss, while high-value CP experiences utility gain. For high values of $c$ ($c = 0.8$), the utility loss for low-value CP is more major, while the utility gain for high-value CP is smaller. 

\subsubsection{Impact of User Elasticity}
To observe how the elasticity of the users impacts the market decisions, we have simulated the market equilibria for $\alpha = 0.2$ and $\alpha = 0.8$, which are lower and higher than the benchmark $\alpha = 0.5$, respectively. 

\begin{figure*}[t]
\centering
\subfigure[ZRE, $\alpha=0.2$]{\includegraphics[width=1.35in, angle=0]{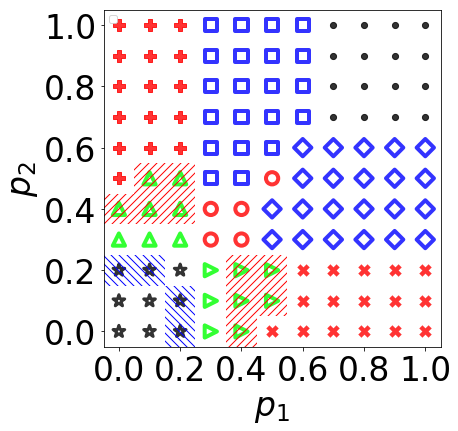}}
\hspace*{0.15in}
\subfigure[CP $1$'s utility changes]{\includegraphics[width=2in, angle=0]{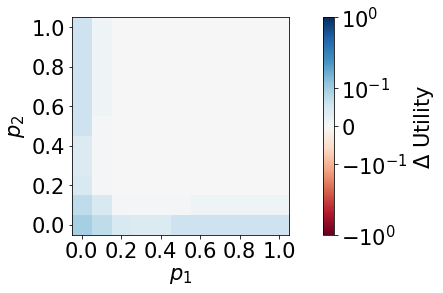}} 
\hspace*{-0.65in}
\subfigure[CP $2$'s utility changes]{\includegraphics[width=2in, angle=0]{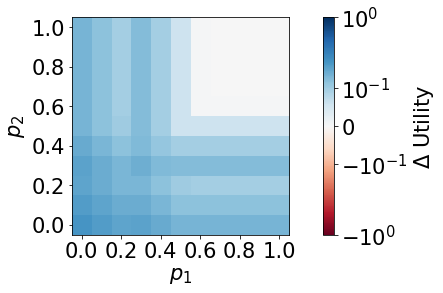}} 
\hspace*{-0.65in}
\subfigure[HHI changes]{\includegraphics[width=2in, angle=0]{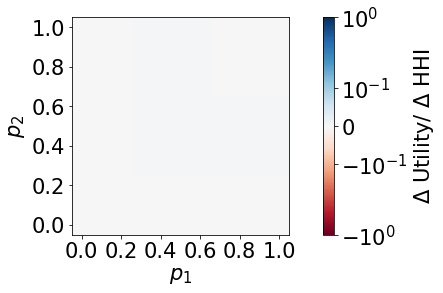}} 

\subfigure[ZRE, $\alpha=0.8$]{\includegraphics[width=1.35in, angle=0]{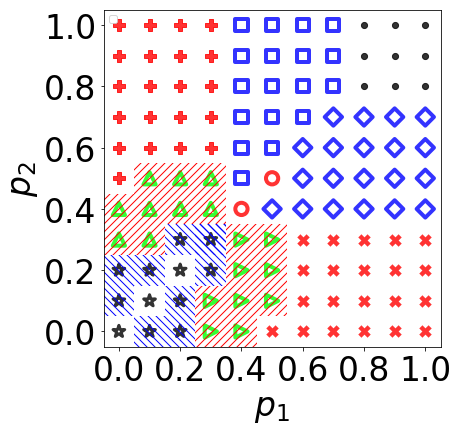}}
\hspace*{0.15in}
\subfigure[CP $1$'s utility changes]{\includegraphics[width=2in, angle=0]{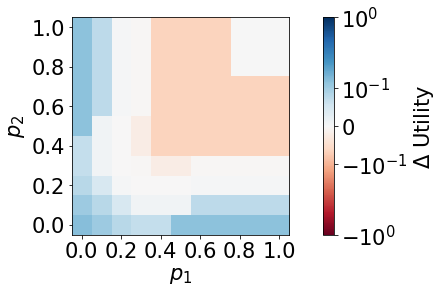}} 
\hspace*{-0.65in}
\subfigure[CP $2$'s utility changes]{\includegraphics[width=2in, angle=0]{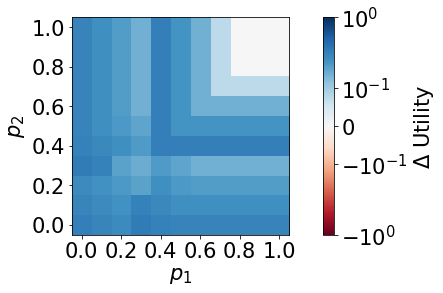}} 
\hspace*{-0.65in}
\subfigure[HHI changes]{\includegraphics[width=2in, angle=0]{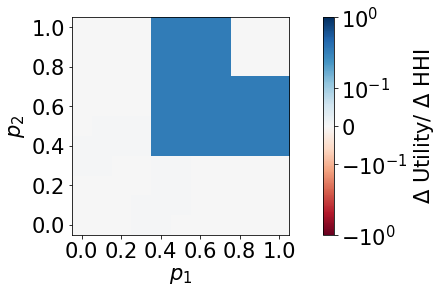}} 
\caption{(a,d) ZRE under complementary duopoly. (b,c,e,f) The differences in CPs' utilities when zero-rating is available and the market reaches equilibria, minus when zero-rating is not available. We have: $c=0.5$, $\boldsymbol{\phi} = (0.1, 0.4, 0.4, 0.1)$, $\boldsymbol{\psi} = (0.2, 0.4, 0.4)$, $\boldsymbol{\delta} = (1.0, 1.0)$ and $\boldsymbol{q}=(0.4,1.0)$.}
\label{fig:impact_alpha}
\end{figure*}

Figure \ref{fig:impact_alpha}(a, e) depicts the impact of $\alpha$ on the equilibria. As the user elasticity increases, the providers are more inclined toward zero-rating since they can attract the elastic users of the market and increase their utilities. We observe that the regions of zero-rating pressure increase as well with the user elasticity, because in some ISP prices where only one CP is willing to zero-rate under any $\Theta$, if the other CP does not it would lose more customers due to high elasticity. 

Figure \ref{fig:impact_alpha} also depicts the corresponding utility and HHI changes corresponding to Figure \ref{fig:impact_alpha}(a, d). We observe that low-value CP $1$ mostly has utility loss as $\alpha$ increases. With higher user elasticity, the number CP $1$'s elastic users increases. Since CP $1$ is less likely to afford zero-rating compared to CP $2$, high-value CP $2$ could attract elastic users if it is the only one who zero-rates. As before, HHI remains increasing with is an indication of the market moving towards a monopoly.

\emph{Takeaways:} 
When user elasticity increases, it mostly harms low-value CP during zero-rating, since the higher fraction of its users would be elastic and would leave it when it does not afford zero-rating.

\subsubsection{Impact of Baseline Market Shares}
To show the impact of baseline market shares, $\boldsymbol{\phi}$ and $\boldsymbol{\psi}$, we vary the values of $\phi_1$ and $\phi_2$ for CP $1$ and CP $2$, and $\psi_1$ and $\psi_2$ for ISP $1$ and ISP $2$. In our benchmark simulation in Figure \ref{fig:com_duopoly} and \ref{fig:CP_utility_baseline}, we showed how ZRE, utilities and HHI look when $\phi_1 = \phi_2$ and $\psi_1 = \psi_2$. In this section, we evaluate our model for $\phi_1 > \phi_2$, $\phi_1 < \phi_2$, $\psi_1 > \psi_2$, and $\psi_1 < \psi_2$.

\begin{figure*}[t]
\centering
\subfigure[ZRE,\newline$\boldsymbol{\phi} = (0.1, 0.2, 0.6, 0.1)$]{\includegraphics[width=1.35in, angle=0]{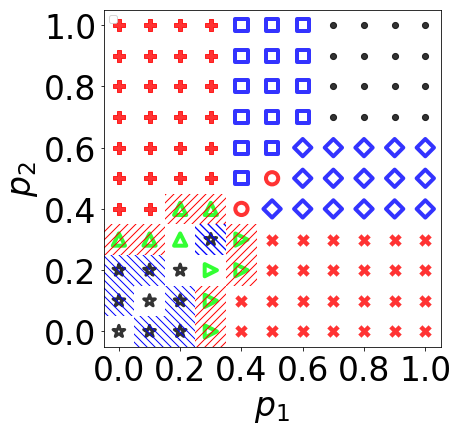}}
\hspace*{0.15in}
\subfigure[CP $1$'s utility changes]{\includegraphics[width=2in, angle=0]{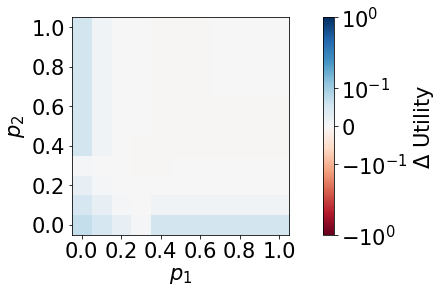}} 
\hspace*{-0.65in}
\subfigure[CP $2$'s utility changes]{\includegraphics[width=2in, angle=0]{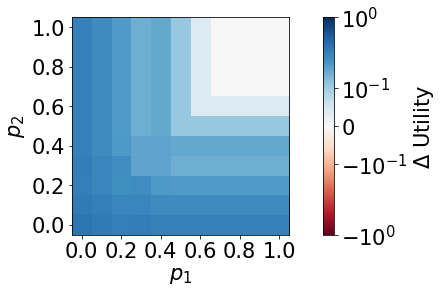}} 
\hspace*{-0.65in}
\subfigure[HHI changes]{\includegraphics[width=2in, angle=0]{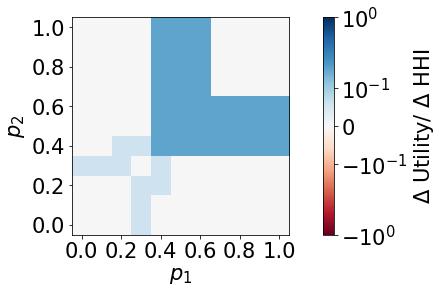}} 

\subfigure[ZRE,\newline $\boldsymbol{\phi} = (0.1, 0.6, 0.2, 0.1)$]{\includegraphics[width=1.35in, angle=0]{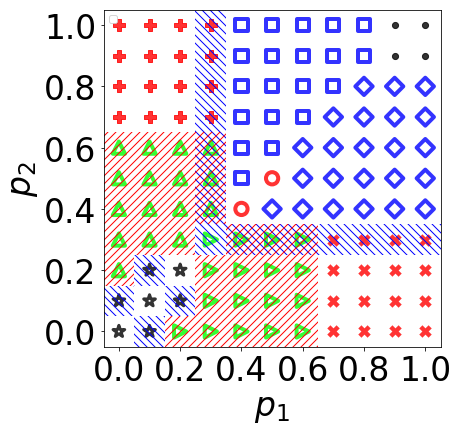}}
\hspace*{0.15in}
\subfigure[CP $1$'s utility changes]{\includegraphics[width=2in, angle=0]{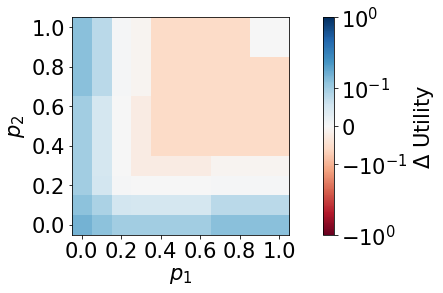}} 
\hspace*{-0.65in}
\subfigure[CP $2$'s utility changes]{\includegraphics[width=2in, angle=0]{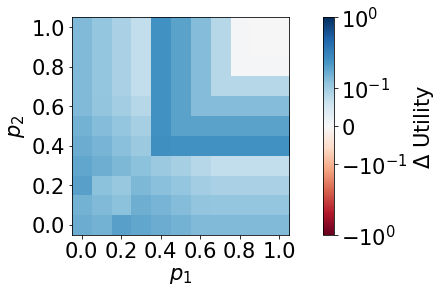}} 
\hspace*{-0.65in}
\subfigure[HHI changes]{\includegraphics[width=2in, angle=0]{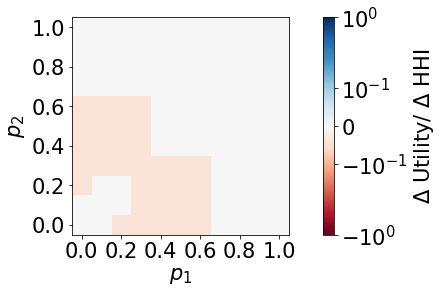}} 
\caption{(a,d) ZRE under complementary duopoly. (b,c,e,f) The differences in CPs' utilities when zero-rating is available and the market reaches equilibria, minus when zero-rating is not available. 
We have: $\alpha = 0.5$, $c=0.5$, $\boldsymbol{\psi} = (0.2, 0.4, 0.4)$, $\boldsymbol{\delta} = (1.0, 1.0)$ and $\boldsymbol{q}=(0.4,1.0)$.}
\label{fig:impact_phi}
\end{figure*}

Figure \ref{fig:impact_phi}(a, e) illustrates the impact of CPs' baseline market shares on the equilibria. We observe that when the baseline market share of the low-value (high-value) CP increases (decreases), the high-value CP is more willing to establish zero-rating, as it wants to increase its small initial market share, and since it has a high value, it can afford that when ISP prices are not too high. 

Figure \ref{fig:impact_phi} also shows the changes of CPs' utilities and the HHI corresponding to \ref{fig:impact_phi}(a, d) when $\boldsymbol{\phi}$ is being varied. We observe that when low-value CP $1$ is also the one with the smaller market share, zero-rating being available causes its utility to sometimes slightly decrease when CP $1$ cannot afford zero-rating while CP $2$ can. Since this decrease is small, it might not be observable in Figure \ref{fig:impact_phi}(b). 

However, when low-value CP $1$ has a higher market share than high-value CP $2$, it also will have more elastic customers. When CP $2$ establishes zero-rating relations it will attract more customers from CP $1$ and as a result, CP $1$ will have a major drop (the red regions in Figure \ref{fig:impact_phi}(f)). The utility of high-value CP $2$ is generally increased compared to the case where no zero-rating is allowed. Note that CP $1$'s utility is higher when its baseline market share is higher, but Figure \ref{fig:impact_phi} merely illustrates the difference in utilities when zero-rating is available minus when it is not. This represents the only scenario where zero-rating can potentially have some benefit in terms of the Herfindahl index, as the introduction of zero-rating increases the market share of the CP with the lower baseline market share. 
Practically, this scenario translates to a case where a CP that is making more profit per customer has a lower initial market share, and it can directly increase its market share by reducing its profitability to move closer to the other CP. However, it is not clear that zero-rating is needed to address the market share imbalance and hence long term consumer benefit since it introduces a utility loss to the low-value CP. In other words, even though the variations in ISPs' baseline market shares change the equilibria pattern, the low-value CP would still usually have utility loss, whereas the high-value CP would have utility gain.

\begin{figure*}[t]
\centering
\subfigure[ZRE,\newline
$\boldsymbol{\psi} = (0.2, 0.2, 0.6)$ ]{\includegraphics[width=1.35in, angle=0]{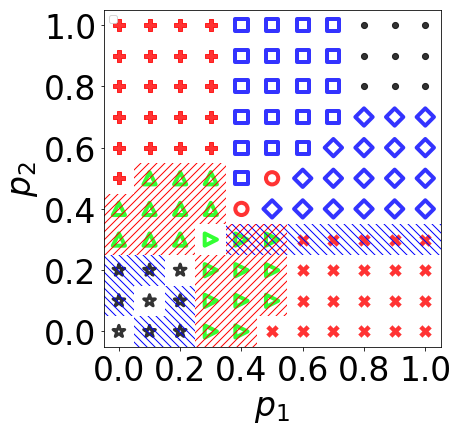}}
\hspace*{0.15in}
\subfigure[CP $1$'s utility changes]{\includegraphics[width=2in, angle=0]{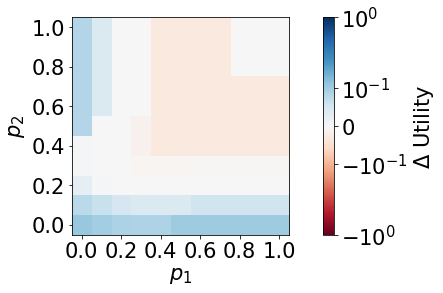}} 
\hspace*{-0.65in} 
\subfigure[CP $2$'s utility changes]{\includegraphics[width=2in, angle=0]{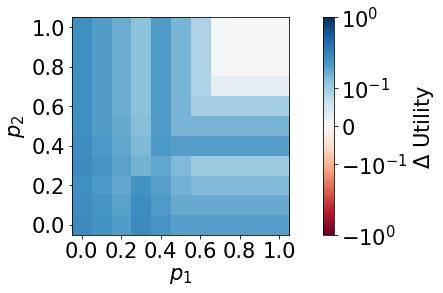}} 
\hspace*{-0.65in} 
\subfigure[HHI changes]{\includegraphics[width=2in, angle=0]{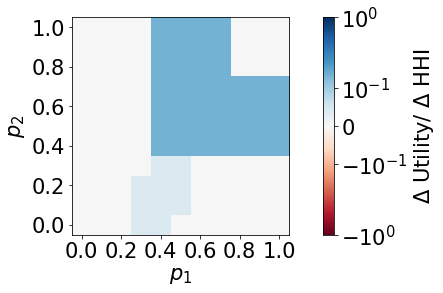}} 

\subfigure[ZRE,\newline 
$\boldsymbol{\psi} = (0.2,0.6,0.2)$]{\includegraphics[width=1.35in, angle=0]{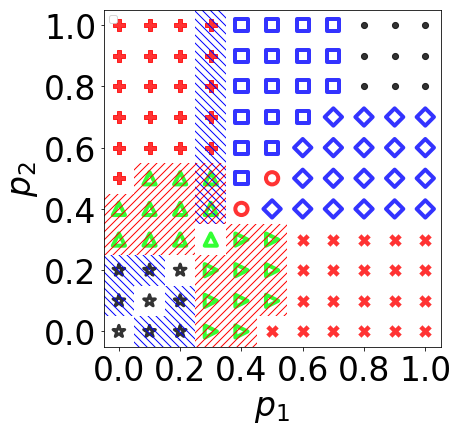}}
\hspace*{0.15in}
\subfigure[CP $1$'s utility changes]{\includegraphics[width=2in, angle=0]{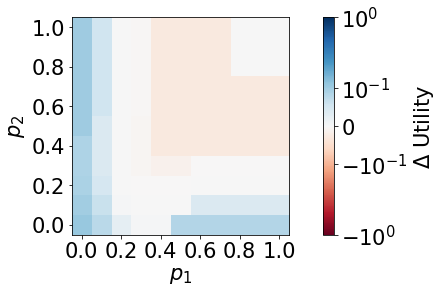}} 
\hspace*{-0.65in} 
\subfigure[CP $2$'s utility changes]{\includegraphics[width=2in, angle=0]{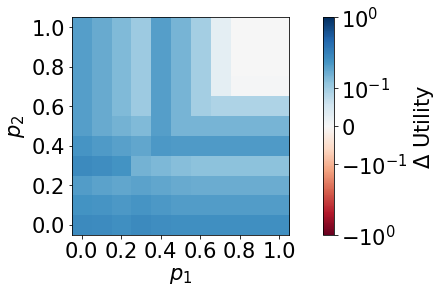}} 
\hspace*{-0.65in} 
\subfigure[HHI changes]{\includegraphics[width=2in, angle=0]{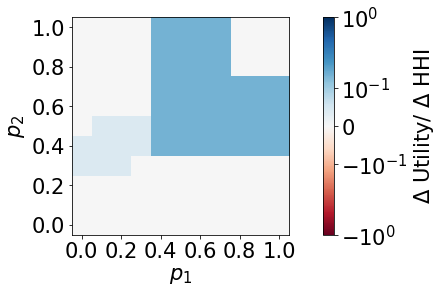}} 

\caption{(a,d) ZRE under complementary duopoly. (b,c,e,f) The differences in CPs' utilities when zero-rating is available and the market reaches equilibria, minus when zero-rating is not available. 
We have: $\alpha = 0.5$, $c=0.5$, $\boldsymbol{\phi} = (0.1, 0.4, 0.4, 0.1)$, $\boldsymbol{\delta} = (1.0, 1.0)$ and $\boldsymbol{q}=(0.4,1.0)$.}
\label{fig:impact_psi}
\end{figure*}

Figure \ref{fig:impact_psi} shows the impact of $\boldsymbol{\psi}$ on the equilibria, CPs' utilities, and HHI. From (a, d), we observe that CPs are generally more willing to zero-rate with the ISP who has a higher baseline market share to attract more customers, and hence asymmetries are introduced to the graphs when $\psi_1 \neq \psi_2$. Therefore, compared to Figure \ref{fig:com_duopoly}, zero-rating pressure happens in higher price ranges of the ISP who has a smaller $\psi$ since CP $1$ is less willing to zero-rate with it without CP $2$ zero-rating. The highest utility drop for CP $1$ mainly happens when CP $2$ establishes more zero-ratings than CP $1$ while CP $2$'s utility generally increases. Furthermore, HHI in both cases are generally increasing.

\emph{Takeaways:} 
As the baseline market share of low-value CP increases (the baseline market share of high-value CP decreases), zero-rating mostly harms low-value CP's utility compared to the case where it is not available. With zero-rating available, high-value CP affords and has more incentives to zero-rate, and may attract elastic customer of low-value CP. As the baseline market shares of two ISPs become asymmetrical, CPs are more willing to zero-rate with the ISP with higher market share, where high-value CP affords to do so more than its low-value competitor, resulting in utility gains for high-value CP at the expense of utility loss for the low-value one. 

Based on the utility analysis in this section, Table \ref{Tab:impact} summarizes how different parameters impact CPs' average utilities and market shares over different ISP prices when zero-rating is available versus when it is not. This average is computed for all the graphs in this section, were the utility and market share values for $121$ points were recorded over $p_1, p_2 \in \{0.0, 0.1, 0.2, ..., 1.0\}$. We see that regardless of parameter choice, aside from some exceptions, zero-rating mostly favors the utility and market share of high-value CP $2$, while it harms the market share of low-value CP $1$. Moreover, even though CP $1$'s average utility mostly increases, its utility drops in a large ISP price range as well, and its gain is not as major as CP $2$'s average utility gain. As we saw in this section, the utility gains for CP $1$ mostly happen in low ISP prices where both CPs afford to zero-rate. However, in higher ISP prices where the competition increases and CP $2$ is mostly the only CP who zero-rates, there is a big range where CP $1$'s utility drops.

\setlength{\tabcolsep}{1.5pt}
\begin{table*}[ht!]
 \caption{~\looseness=-1 Impact of parameters on CPs' average utilities after ZER compared to when zero-rating is not available. The average is taken over different ISP prices $p_1, p_2 \in \{0, 0.1,..., 1.0\}$. Benchmark parameters are $\alpha = c = 0.5$, $\boldsymbol{\phi} = (0.1, 0.4, 0.4, 0.1)$, $\boldsymbol \psi = (0.2, 0.4,0.4)$, $\boldsymbol{\delta} = (1.0, 1.0)$ and $\boldsymbol{q}=(0.4,1.0)$.} \centering
 \footnotesize
 \begin{tabular}{cccccc}
 \toprule
 \multicolumn{2}{c}{\multirow{2}{*}{parameter changes against benchmark}}& \multicolumn{2}{c}{\ \ \ low-value CP's avg of\ \ \ } & \multicolumn{2}{c}{\ \ \ high-value CP's avg of\ \ \ } \\
 \cmidrule(lr){3-4} \cmidrule(lr){5-6} 
 & & utility & market share & utility & market share \\
 \midrule
 benchmark parameters &---& \textcolor{green}{$\uparrow$} & \textcolor{red}{$\downarrow$} & \textcolor{green}{$\uparrow$} & \textcolor{green}{$\uparrow$} \\
 \midrule
 \multirow{2}{*}{bandwidth usage coefficient} & decrease ($c=0.2$) & \textcolor{green}{$\uparrow$} & \textcolor{red}{$\downarrow$} & \textcolor{green}{$\uparrow$} & \textcolor{green}{$\uparrow$} \\
 & increase ($c=0.8$) & \textcolor{red}{$\downarrow$} & \textcolor{red}{$\downarrow$} & \textcolor{green}{$\uparrow$} & \textcolor{green}{$\uparrow$} \\
 \midrule
 \multirow{2}{*}{user elasticity} & decrease ($\alpha=0.2$) & \textcolor{green}{$\uparrow$} & \textcolor{red}{$\downarrow$} & \textcolor{green}{$\uparrow$} & \textcolor{green}{$\uparrow$} \\
 & increase ($\alpha=0.8$) & \textcolor{green}{$\uparrow$} & \textcolor{red}{$\downarrow$} & \textcolor{green}{$\uparrow$} & \textcolor{green}{$\uparrow$} \\
 \midrule
 \multirow{4}{*}{CPs' baseline market shares} & decrease low-value CP's baseline & \textcolor{green}{$\uparrow$} & \textcolor{red}{$\downarrow$} & \textcolor{green}{$\uparrow$} & \textcolor{green}{$\uparrow$} \\
 & ($\phi = (0.1,0.2,0.6,0.1)$) & & & & \\
 & increase low-value CP's baseline & \textcolor{green}{$\uparrow$} & \textcolor{red}{$\downarrow$} & \textcolor{green}{$\uparrow$} & \textcolor{green}{$\uparrow$} \\
 & ($\phi = (0.1,0.6,0.2,0.1)$) & & & & \\
 \midrule
 \multirow{4}{*}{ISPs' baseline market shares} & decrease ISP $1$'s baseline & \textcolor{green}{$\uparrow$} & \textcolor{red}{$\downarrow$} & \textcolor{green}{$\uparrow$} & \textcolor{green}{$\uparrow$} \\
 & ($\psi = (0.2,0.2,0.6)$) & & & & \\
 & increase ISP $1$'s baseline & \textcolor{green}{$\uparrow$} & \textcolor{red}{$\downarrow$} & \textcolor{green}{$\uparrow$} & \textcolor{green}{$\uparrow$} \\
 & ($\psi = (0.2,0.6,0.2)$) & & & & \\
 \bottomrule
 \end{tabular}
 \label{Tab:impact}
 \end{table*}

\section{Related Work}\label{sec:rel-work}
Cheng et al. \cite{cheng2011debate} and Ma \cite{Ma17pay} consider the case where CPs bargain with the monopolistic ISP to obtain exclusive priority for their traffic; CPs are charged a fee only if they opt for priority, and users can access one content provider exclusively. 
While they both define a fixed market share for CPs, Cheng et al. incorporate consumer surplus for a case of monopolistic ISP and find that premium peering leaves content providers worse off, but Ma assumes ISPs are always willing to offer exclusive priorities, while CPs are the decision-makers. In our work, 
we consider zero-rating decisions in the market of multiple CPs and ISPs and study the case where customers do not necessarily use exclusive CPs. We consider both CPs' and ISPs' zero-rating decisions 
and show that zero-rating may cause market distortion by increasing the Herfindahl index in the market of CPs, and usually leaves the low-value CP (startups with low incomes) worse off.

Choi and Kim \cite{pil2010net} analyze the effects of net neutrality regulations on investment incentives for ISPs and CPs
. Reggiani et al. \cite{reggiani2016net} also model an Internet broadband provider that can offer a priority to two different content providers, low-value and high-value, and show that net neutrality regulations effectively protect innovation done at the edge by small content providers. Zhang et al. \cite{zhang2015sponsored} builds a two-class service model to analyze the
consumers' traffic demand under the sponsored data plan
with consideration of QoS, and characterize the interaction between CPs and a monopolistic ISP. 
Wong et al. Shirmali \cite{shrimali2008surplus} considers surplus extraction by a monopolistic ISP who controls the medium of information transfer between application developers and consumers, and shows that net neutrality is necessary to ensure maximal benefit to the society. Wong et al. \cite{joe2018sponsoring} formulate an analytical model of the user, CP, and ISP interactions and derives their optimal
behaviors. They show that zero-rating
disproportionately benefits less cost-sensitive CPs and more cost-sensitive users, exacerbating disparities among CPs
. While the aforementioned models consider a market of monopolistic ISP and duopolistic CPs in which users access exclusively one content provider, our model extends to a larger market of ISPs and CPs where users are not required to access content providers exclusively.

Some previous studies focus on abolishing net neutrality under zero-rating. For instance, the authors in \cite{somogyi2017economics,jeitschko2017zero,kies2017impacts} analyze zero-rating incentives of a monopolistic ISP in a homogeneous market of customers, and how different zero-rating equilibria impacts social welfare. Somogyi et al. \cite{somogyi2017economics} consider capacity constraints in a two-sided market with advertising revenue for CPs. Jullien et al. \cite{jullien2018internet} discuss the elasticity of users and mainly focus on the case of a monopoly network with inelastic participation of consumers. While they all focus on a small monopolistic ISP market, we extend our study to larger markets 
where our main focus is on how zero-rating impacts innovations in the market and we do not analyze the customers' side in detail.

Some other work study real-word markets which have established zero-rating. Mathur et al. \cite{mathur2015mixed} analyze network usage data in South Africa and show that where usage-based billing is prevalent and data costs are high, users are extremely cost-conscious and are willing to use ISPs/CPs who have more affordable costs. Using the Android app MySpeedTest \cite{chetty2013measuring}, they collect information on mobile data usage 
and find that 90\% of users consumed twice as much data when they do not pay for ISP bandwidth compared to when they have a usage-based mobile connection. Chen et al. \cite{chen2017exploring} also collect a dataset by MySpeedTest 
and by analyzing zero-rating WhatsApp on Cell-C's network and zero-rating twitter on MTN's network, they find that zero-rating increases overall usage of the WhatsApp on Cell-C and Twitter on MTN network while it decreases it on most other providers. 
Valancius et al. \cite{valancius2011many} also innovatively use existing data to determine some 
unknown parameters such as the cost of various resources, or how users respond to price. While in our work we use synthetic parameters to test our model, our final results and takeaways are not qualitatively impacted by parameter choice, albeit our model is flexible to use real-world parameters as a future direction.

\section{Discussions and conclusions}\label{sec:con}
This paper explores a controversial and unsettled aspect of net neutrality by analyzing zero-rating decisions in a market of multiple CPs and ISPs, and their impact on growing businesses and incumbents. We model the zero-rating decisions as a bargain between CPs and ISPs and find the \emph{zero-rating equilibria} resulted in the market. By mainly focusing on CP's side of the market and using the Herfindahl index, we have theoretically shown the distortions in the market may increase when zero-rating is available. We further numerically and qualitatively analyze the impact of zero-rating on CPs with different values and show that zero-rating typically disadvantages low-value CPs and could stunt the innovations. Our results strongly suggest that zero-rating can be harmful to competitiveness in a market, especially when the players have asymmetric market power and hence it should be disallowed under net neutrality.

We have taken into account the decision of ISPs to enhance their market share, where there might be different factors that impact ISPs' revenue differently. First, compared to the data prices ISPs charge users, they might offer a price discount to CPs in case of zero-rating. Second, the bandwidth usage for zero-rated data could be potentially more than that of paid data. We consider the price of ISPs and the value of CPs as the units of their utilities, and by assuming that customers are probabilistically homogeneous, we analyze the market at both a macroscopic and a microscopic level and depend on discrete choice modeling. A future direction would be considering different real-world markets that have been studied in the past and tuning our parameters based on them.

\bibliographystyle{unsrt} 
\bibliography{sample-bibliography}


\section*{Appendix I}\label{appendix:proof}
\noindent {\bf Proof of Lemma \ref{lemma:additivity}:}
The additivity property holds because the users choose an ISP-CP pair based on the linear proportional form of Assumption \ref{assumption:choice}, which induces the market shares of the providers in terms of linear functions in Equation \ref{equation:market_share}. When zero-rating profiles, i.e., $\boldsymbol{\theta_i}$ or $\boldsymbol{\vartheta_j}$ are similar, the closed-form market share expressions keep the same form in terms of $\rho_{ij}(\Theta)$, while the baseline market shares of the providers, i.e., $\phi_{i}$ or $\psi_{j}$, are aggregated.
\done

\vspace{0.1in}

{\bf Proof of Lemma \ref{lemma:high_value}:}
If the content provider with the lower value zero-rates, zero-rating must be a strategy to cause its utility to increase. In that case, based on Equation \ref{equation:utilities}, zero-rating would increase the utility of the high-value CP as well since every characteristic of them are similar, only the high-value one has a higher $q$. In other words, if the low-value CP can afford zero-rating, so does the high-value CP. However, the opposite is not always true; if the high-value CP can afford zero-rating, the low-value CP would not necessarily afford it. Therefore, in the market of two CPs, possible zero-rating equilibria are either only the high-value CP zero-rate, or both zero-rate, or none zero-rates. 
\done

\vspace{0.1in}

{\bf Proof of Lemma \ref{lemma:HHI}:} 
Suppose $X_i$ is the fraction of users (market share) for CP $i$, and is equivalent to $\sum_{j\in \mathcal(M)}\rho_{ij}$ when the total market size $X$ is normalized to $1$. We have
$$HHI = \sum_{i\in \mathbb{N}} X_i ^2 $$
and
$$\sigma^2 = \frac{\sum_{i\in \mathbb{N}}X_i^2}{|\mathbb{N}|}-\mu^2$$
where the mean $\mu = \frac{\sum_{i\in \mathbb{N}}X_i}{|\mathbb{N}|} = \frac{1}{|\mathbb{N}|}$

Therefore, we have:
$$HHI = \frac{1}{|\mathbb{N}|} + \sigma^2|\mathbb{N}|$$
Hence, the Herfindahl index has the minimum value of $\frac{1}{|\mathbb{N}|}$ when the variance of market shares is zero, and it increases as the variance increases. In a market of two content providers, this index grows when the gap between their market shares increases. 
\done

\vspace{0.1in}

{\bf Proof of Lemma \ref{lemma:HHI_ZR_noZR}:} Without loss of generality, suppose we have two actual content providers in the system. Therefore, we will have $2^2$ actual, dummy and auxiliary content providers in our computations. Let's call the actual content providers CP $1$ and CP $2$, the dummy content provider CP $0$, and the combination of CP $1$ and CP $2$ is called CP $3$. We normalize the market size $X$ to $1$. When no zero-rating exists in the system, since $\Theta = \boldsymbol{0}$, based on Equation \ref{equation:market_share}, we have: 
$$X_{1j} = \phi_1\psi_j, \ X_{2j} = \phi_2\psi_j, \ X_{3j} = \phi_3\psi_j$$
Therefore, give Equation \ref{equation:effective_user}:
$$X_1 = \sum_j X_{1j} + X_{3j} =\sum_j \psi_j (\phi_1 + \phi_3) = \phi_1 + \phi_3$$
and
$$X_2 =\sum_j X_{2j} + X_{3j} = \sum_j \psi_j (\phi_2 + \phi_3)=\phi_2 + \phi_3$$
Hence, the Herfindahl index of the actual CPs would be:
$$HHI = \frac{X_1^2+X_2^2}{(X_1+X_2)^2} = \frac{(\phi_1 + \phi_3)^2+(\phi_2 + \phi_3)^2}{(\phi_1+\phi_2 + 2\phi_3)^2}$$

$$\frac{\phi_i\psi_j \theta_{ij}}{\sum_{i'}\sum_{j'}\phi_{i'}\psi_{j'}\theta_{i'j'}}\alpha + \phi_i\psi_j(1-\alpha)$$

On the other hand, if everyone zero-rates in the system, we have: 
$$X_{1j} = \phi_1\psi_j(1-\alpha)+ \frac{\phi_1\psi_j}{\sum_{i'}\sum_{j'}\phi_{i'}\psi_{j'}\theta_{i'j'}}\alpha =
\phi_1\psi_j(1-\alpha)$$

Let's define
$$A =1- \alpha + \frac{\alpha}{\sum_{i'}\sum_{j'}\phi_{i'}\psi_{j'}\theta_{i'j'}}$$
then we have: 
$$X_{1j} =A\times \phi_1\psi_j$$
Similarly, for $X_{2j}$ and $X_{3j}$, we have:
$$ \ X_{2j} = A\times \phi_2\psi_j, \ X_{3j} = A\times\phi_3\psi_j$$
Therefore: 
$$X_1 = \sum_j X_{1j} + X_{3j} =\sum_j\psi_j (A\phi_1 + A\phi_3)=A(\phi_1 + \phi_3)$$
and
$$X_2 =\sum_j X_{2j} + X_{3j} =\sum_j\psi_j (A\phi_2 + A\phi_3)=A(\phi_2 + \phi_3)$$
Hence, the Herfindahl index of the actual CPs would be:
\begin{multline*}
HHI = \frac{X_1^2+X_2^2}{(X_1+X_2)^2} =  \\ \frac{A^2(\phi_1 + \phi_3)^2+A^2(\phi_2 + \phi_3)^2}{A^2(\phi_1+\phi_2 + 2\phi_3)^2} = \frac{(\phi_1 + \phi_3)^2+(\phi_2 + \phi_3)^2}{(\phi_1+\phi_2 + 2\phi_3)^2} 
\end{multline*}
Which is the same Herfindahl index of the case where no zero-rating is allowed in the system. 
\done

\vspace{0.1in}
{\bf Proof of Theorem \ref{theorem:HHI_q_phi}:} Based on Corollary \ref{lemma:high_value}, for any two CPs with different values, the possible zero-rating strategies in the equilibria are either the high-value content provider zero-rates, or both of them zero-rate, or neither of them zero-rate. In the last two cases, the Herfindahl index would be the same according to Lemma \ref{lemma:HHI_ZR_noZR}. Further, when no one zero-rates, the market shares would be the same as when no zero-rating is allowed in the system, therefore the Herfindahl index would be the same. In the first case, based on Corollary \ref{lemma:customers}, the high-value content provider would attract more customers if it zero-rates. Since the high-value content provider is the one with the higher baseline market share, after it zero-rates, its market share would further increase from its baseline. Therefore, the gap between market shares increases as well. This could extend to multiple CPs where the variance of the market shares would increase in this case. Hence, based on Lemma \ref{lemma:HHI}, the Herfindahl index increases as well. 
\done

\vspace{0.1in}

{\bf Proof of Theorem \ref{theorem:high_value_CP_utility}} 
when a low-value CP cannot zero-rate, any zero-rating relation that its high-value opponent establishes will decrease its utility. The reason is that based on Equation \ref{equation:market_share}, in case there is any zero-rating relations in the market ($\Theta\neq \boldsymbol{0}$), the market share of a pair of CP $i$ and ISP $j$ is computed from 
\[\rho_{ij} (\Theta) =\frac{\phi_i\psi_j \theta_{ij}}{\sum_{i'}\sum_{j'}\phi_{i'}\psi_{j'}\theta_{i'j'}}\alpha + \phi_i\psi_j(1-\alpha) \] 

If low-value CP $1$ does not zero-rate, $\theta_{1j} = 0\: \forall j\in \mathbb{M}$. Therefore, the first term in Equation \ref{equation:market_share} will be zero and as a result, if $\theta_{2j} = 1\: \exists j\in \mathbb{M}$, the elastic users of CP $1$ will move to CP $2$. Based on Equation \ref{equation:utilities}, the utility of CP $1$ produced from each ISP $j$ will be computed from the term $U_1^j(\Theta) = q_1 \mathbb{X}_{1j}(\Theta).c$, and as $\mathbb{X}_{ij} = X\rho_{ij}$ decreases, CP $1$'s utility decreases as well. 

However, if CP $2$ zero-rates with ISP $j$, the same analysis cannot be applied to its utility. Even though based on Corollary \ref{lemma:customers}, in case $\theta_{2j} = 1\ \exists j\in \mathbb{M}$ the market share of CP $2$ increases, as it is also paying the bandwidth price of $p_j$ to ISP $j$, 
its utility computation is different than when it is not zero-rating ($U_2^j(\Theta) = (q_2-\delta_jp_j) \mathbb{X}_{2j}(\Theta)$ if $\theta_{2j} = 1$ versus $U_2^j(\Theta) = (q_2) \mathbb{X}_{2j}(\Theta).c$ if $\theta_{2j} = 0$). However, based on the definition of ZRE we can prove its utility would not decrease. In ZRE, if low-value CP does not zero-rate while the high-value CP does, i.e., $\theta_{1j} = 0$ and $\theta_{2j} = 1$, it would not be a ZRE if it decreases high-value CP's utility compared to $\theta_{1j} = 0$ and $\theta_{2j} = 0$. Therefore, when low-value CP $1$ does not zero-rate, any zero-rating relation of CP $2$ after ZRE will either increase its utility or keep it unchanged. 
\done

\vspace{0.1in}



\section*{Appendix II}\label{appendix:delta}
In this section, we consider the case where ISPs decide what price discounts ($\boldsymbol{\delta}$) to offer for zero-rating. These discounts are in the spirit of bulk discounts, as the CP would be paying for {\emph{all}} of its users in bulk. For simplicity of computations, we assume each ISP $j$ chooses $\delta_j$ from the set $\{0.0, 0.1,..., 1.0 \}$. 
We define $\boldsymbol{\bar\delta_j} \triangleq \{0.0, 0.1,...,1.0\} - \{\delta_j\}$, and we denote the discount strategy of all ISPs except for ISP $j$ by $\boldsymbol \delta_{-j}$. A discount profile $\boldsymbol\delta$ is defined as a \emph{Nash equilibrium} if there exists a ZRE built on that, and for the resulting ISP revenues we have $R_j|_{(\delta_j, \boldsymbol\delta_{-j})}\geq R_j|_{ (\boldsymbol{\bar\delta_j}, \boldsymbol\delta_{-j})}$ for all ISP $j$. Since there might be multiple values of $\boldsymbol\delta$ satisfying this condition, we choose $\boldsymbol\delta$ to be the maximum among those who lead to equilibria. We further assume that the more expensive ISP is the tie breaker. For instance, if ISP $2$ has a higher per bandwidth unit price than ISP $1$, i.e., $p_2 \geq p_1$, in case we have a tie between different vectors of $\boldsymbol\delta$, the one will be chosen which has a higher $\delta_2$.

\begin{figure*}[t]
\centering
\subfigure[ZRE, varying $\boldsymbol{\delta}$ ]{\includegraphics[width=1.35in, angle=0]{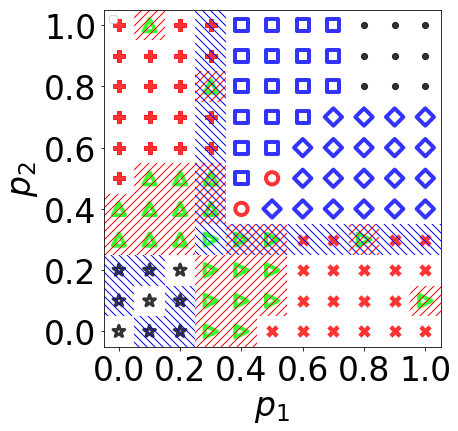} }
\hspace*{0.15in} 
\subfigure[CP $1$'s utility changes ]{\includegraphics[width=2in, angle=0]{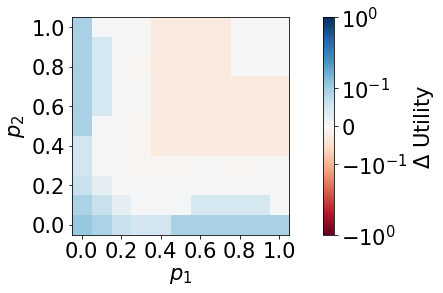}} 
\hspace*{-0.65in} 
\subfigure[CP $2$'s utility changes]{\includegraphics[width=2in, angle=0]{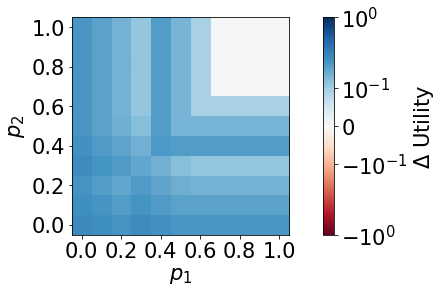}}
\hspace*{-0.65in} 
\subfigure[HHI changes]{\includegraphics[width=2in, angle=0]{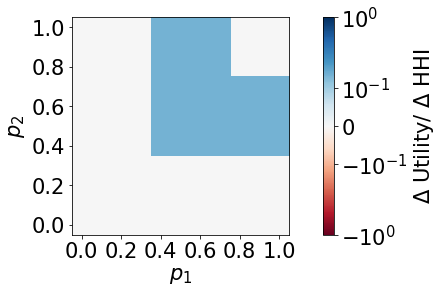}} 
\caption{zero-rating equilibria and the differences in CPs' utilities and HHI when zero-rating is available and the market reaches equilibria, minus when zero-rating is not available. We have: $\alpha=0.5$, $c=0.5$, $\boldsymbol{\phi} = (0.1, 0.4, 0.4, 0.1)$, $\boldsymbol{\psi} = (0.2, 0.4, 0.4)$, and $\boldsymbol{q}=(0.4,1.0)$.}
\label{fig:CP_utility_delta}
\end{figure*}

We again use our benchmark scenario on Figure \ref{fig:com_duopoly} where we had $\boldsymbol{q} = (0.4,1.0)$, $\boldsymbol \phi = (0.1, 0.4, 0.4, 0.1)$, $\boldsymbol \psi = (0.2,0.4,0.4)$, $\alpha = 0.5$, and $c = 0.5$. Figure \ref{fig:CP_utility_delta} (a) depicts the final equilibria in the market where both CPs and ISPs make zero-rating decisions, and Figure \ref{fig:CP_utility_delta} depicts CPs' utilities and HHI changes corresponding to Figure \ref{fig:CP_utility_delta} (a). Furthermore, Table \ref{Tab:discount5_1} shows the discount profiles $\boldsymbol{\delta}$ of the ISPs. Compared to Figure \ref{fig:com_duopoly} where $\boldsymbol{\delta} = \boldsymbol{1}$ and it is remained constant, we observe that the number of zero-rating relations increases after ZRE. The reason is that as Table \ref{Tab:discount5_1} shows, the ISPs may offer discounts for some price profiles so that they attract CPs to zero-rate with them, and as a result expand their market. Note that these price discounts are derived from Nash equilibria and as ISPs are also decision maker agents, they choose their discounts high enough to maximize their rewards given a particular profile the rest of the market has.

\begin{table*}[tb!]
\centering
\footnotesize{
\begin{tabular}{c| c| c| c| c| c| c| c| c| c| c |c} 
\backslashbox{$p_2$}{$p_1$} & $ 0.$ & $ 0.1$ & $ 0.2$ & $ 0.3$ & $ 0.4$ & $ 0.5$ & $ 0.6$ & $ 0.7$ & $ 0.8$ & $ 0.9$ & $ 1.0$ \\
\hline
$ 0.0 $&($ 1.0 , 1.0 $)&($ 1.0 , 1.0 $)&($ 1.0 , 1.0 $)&($ 1.0 , 1.0 $)&($ 1.0 , 1.0 $)&($ 1.0 , 1.0 $)&($ 1.0 , 0.9 $)&($ 1.0 , 0.8 $)&($ 1.0 , 0.7 $)&($ 1.0 , 0.6 $)&($ 1.0 , 0.5 $)\\ 
 \hline

$ 0.1 $&($ 1.0 , 1.0 $)&($ 1.0 , 1.0 $)&($ 1.0 , 1.0 $)&($ 1.0 , 1.0 $)&($ 1.0 , 1.0 $)&($ 1.0 , 1.0 $)&($ 1.0 , 0.9 $)&($ 1.0 , 0.8 $)&($ 1.0 , 0.7 $)&($ 1.0 , 0.6 $)&($ 1.0 , 0.5 $)\\ 
 \hline

$ 0.2 $&($ 1.0 , 1.0 $)&($ 1.0 , 1.0 $)&($ 1.0 , 1.0 $)&($ 1.0 , 1.0 $)&($ 1.0 , 1.0 $)&($ 1.0 , 1.0 $)&($ 1.0 , 1.0 $)&($ 1.0 , 0.8 $)&($ 1.0 , 0.7 $)&($ 1.0 , 0.6 $)&($ 1.0 , 1.0 $)\\ 
 \hline

$ 0.3 $&($ 1.0 , 1.0 $)&($ 1.0 , 1.0 $)&($ 1.0 , 1.0 $)&($ 1.0 , 1.0 $)&($ 1.0 , 1.0 $)&($ 1.0 , 1.0 $)&($ 1.0 , 1.0 $)&($ 1.0 , 1.0 $)&($ 1.0 , 0.7 $)&($ 1.0 , 1.0 $)&($ 1.0 , 0.6 $)\\ 
 \hline

$ 0.4 $&($ 1.0 , 1.0 $)&($ 1.0 , 1.0 $)&($ 1.0 , 1.0 $)&($ 1.0 , 1.0 $)&($ 1.0 , 1.0 $)&($ 1.0 , 1.0 $)&($ 1.0 , 1.0 $)&($ 1.0 , 1.0 $)&($ 1.0 , 1.0 $)&($ 1.0 , 1.0 $)&($ 1.0 , 0.6 $)\\ 
 \hline

$ 0.5 $&($ 1.0 , 1.0 $)&($ 1.0 , 1.0 $)&($ 1.0 , 1.0 $)&($ 1.0 , 1.0 $)&($ 1.0 , 1.0 $)&($ 1.0 , 1.0 $)&($ 1.0 , 1.0 $)&($ 1.0 , 0.9 $)&($ 1.0 , 0.8 $)&($ 1.0 , 0.7 $)&($ 1.0 , 0.6 $)\\ 
 \hline

$ 0.6 $&($ 0.9 , 1.0 $)&($ 0.9 , 1.0 $)&($ 1.0 , 1.0 $)&($ 1.0 , 1.0 $)&($ 1.0 , 1.0 $)&($ 1.0 , 1.0 $)&($ 1.0 , 1.0 $)&($ 1.0 , 1.0 $)&($ 1.0 , 0.8 $)&($ 1.0 , 0.7 $)&($ 1.0 , 1.0 $)\\ 
 \hline

$ 0.7 $&($ 0.8 , 1.0 $)&($ 0.8 , 1.0 $)&($ 0.8 , 1.0 $)&($ 1.0 , 1.0 $)&($ 1.0 , 1.0 $)&($ 0.9 , 1.0 $)&($ 1.0 , 1.0 $)&($ 1.0 , 1.0 $)&($ 1.0 , 1.0 $)&($ 1.0 , 1.0 $)&($ 1.0 , 1.0 $)\\ 
 \hline

$ 0.8 $&($ 0.7 , 1.0 $)&($ 0.7 , 1.0 $)&($ 0.7 , 1.0 $)&($ 0.7 , 1.0 $)&($ 1.0 , 1.0 $)&($ 0.8 , 1.0 $)&($ 0.8 , 1.0 $)&($ 1.0 , 1.0 $)&($ 1.0 , 1.0 $)&($ 1.0 , 0.9 $)&($ 1.0 , 0.8 $)\\ 
 \hline

$ 0.9 $&($ 0.6 , 1.0 $)&($ 0.6 , 1.0 $)&($ 0.6 , 1.0 $)&($ 1.0 , 1.0 $)&($ 1.0 , 1.0 $)&($ 0.7 , 1.0 $)&($ 0.7 , 1.0 $)&($ 1.0 , 1.0 $)&($ 0.9 , 1.0 $)&($ 0.9 , 0.9 $)&($ 0.9 , 0.8 $)\\ 
 \hline

$ 1.0 $&($ 0.5 , 1.0 $)&($ 0.5 , 1.0 $)&($ 1.0 , 1.0 $)&($ 0.6 , 1.0 $)&($ 0.6 , 1.0 $)&($ 0.6 , 1.0 $)&($ 1.0 , 1.0 $)&($ 1.0 , 1.0 $)&($ 0.8 , 1.0 $)&($ 0.8 , 0.9 $)&($ 0.8 , 0.8 $)\\ 
 \bottomrule
\end{tabular}
}
\caption{Discount profile $\boldsymbol\delta$ of ISPs under complementary duopoly ZRE of Figure \ref{fig:CP_utility_delta}.}
\label{Tab:discount5_1}
 \end{table*}

As in Figure \ref{fig:CP_utility_delta}, we observe the difference in utilities of CPs, and market HHI after equilibria when zero-rating is available minus when it is not. We see that the general pattern of the utility loss for low-value CP $1$ and the utility gain for high-value CP $2$ are similar to the benchmark in Figure \ref{fig:CP_utility_baseline}, with slight changes due to ISP price discounts. The reason for these changes is also that offering a discount would have a similar impact on CPs' side of the market as if the ISP has a cheaper price. 

We conclude that even when ISPs make decisions on what price discounts to offer, the general behavior of the market and the notion that zero-rating harming the low-value CP and favoring the high-value one remains unchanged, but we have included this analysis in the Appendix for completeness. Note that since the graphs for different parameter selections are also qualitatively similar to those in Section \ref{sec:eval} and mostly follow the same trends, we have not included them to the Appendix.

\end{document}